\documentclass[12pt,a4paper]{article}

\usepackage{graphicx}

\usepackage{cite}

\usepackage{amsmath}
\usepackage{amssymb}

\newenvironment{Itemize}{\begin{list}{$\bullet$}%
{\setlength{\topsep}{0.2mm}\setlength{\partopsep}{0.2mm}%
\setlength{\itemsep}{0.2mm}\setlength{\parsep}{0.2mm}}}%
{\end{list}}
\newcounter{enumct}
\newenvironment{Enumerate}{\begin{list}{\arabic{enumct}.}%
{\usecounter{enumct}\setlength{\topsep}{0.2mm}%
\setlength{\partopsep}{0.2mm}\setlength{\itemsep}{0.2mm}%
\setlength{\parsep}{0.2mm}}}{\end{list}}

\newlength{\abstwidth}
\begin{document}
\sloppy

\pagestyle{empty}

\begin{flushright}
LU TP 17-22\\
MCnet-17-10\\
June 2017
\end{flushright}

\vspace{\fill}

\begin{center}
{\LARGE\bf The Development of MPI Modelling\\[3mm] in PYTHIA%
\footnote{To be published in "Multiple Parton Interactions at the LHC",
P. Bartalini and J. R. Gaunt, eds., World Scientific}}\\[10mm]
{\Large Torbj\"orn Sj\"ostrand}\\[3mm]
{\it Theoretical Particle Physics,
Department of Astronomy and Theoretical Physics,}\\[1mm]
{\it Lund University, SE-223 62 Lund, Sweden}\\[2mm]
\end{center}

\vspace{\fill}

\begin{center}
\begin{minipage}{\abstwidth}
{\bf Abstract}\\[2mm]
Many of the basic ideas in multiparton interaction (MPI) phenomenology
were first developed in the context of the \textsc{Pythia} event
generator, and MPIs have been central in its modelling of both
minimum-bias and underlying-event physics in one unified framework.
This chapter traces the evolution towards an increasingly sophisticated
description of MPIs in \textsc{Pythia}, including topics such as
the ordering of MPIs, the regularization of the divergent QCD cross
section, the impact-parameter picture, colour reconnection,
multiparton PDFs and beam remnants, interleaved and intertwined
evolution, and diffraction.
\end{minipage}
\end{center}

\vspace{\fill}


\clearpage

\pagestyle{plain}
\setcounter{page}{1}

\section{Introduction}
\label{tssec:intro}

The \textsc{Pythia} event generator \cite{Bengtsson:1982jr} was
initially created to explore the physics of colour flow in hadronic
collisions, in analogy with how the Lund string model
\cite{Andersson:1983ia} had successfully predicted string effects in
$e^+e^-$ annihilation \cite{Andersson:1980vk,Bartel:1981kh}.
Initially only $2 \to 2$ partonic ($q, g, \gamma$)  processes were
implemented, with colour flow connecting the scattered partons to the
beam remnants, followed by string fragmentation using \textsc{Jetset}
\cite{Sjostrand:1982fn}. At the 1984 Snowmass workshop on the SSC,
when I first got directly involved in the physics of high-energy
hadron colliders, it was obvious that this approach was too primitive
to be of relevance. During the autumn I implemented initial- and
final-state radiation (ISR and FSR) \cite{Sjostrand:1985xi}, with the
expectation that this further activity would give event topologies more
comparable with S$p\overline{p}$S data. In terms of jet phenomenology
it did, but underlying events were still much less active than in data.

The natural explanation, in my opinion, was that the composite nature
of the proton would lead to several parton--parton interactions,
giving more activity. Thus in the spring of 1985 I developed a first
multiparton interaction (MPI) model, still primitive but offering a
significantly improved description of data, convincing me that MPIs
was the way to go. Not everybody approved; the first writeup
\cite{Sjostrand:1985vv} was not accepted for publication.
In 1986 studies resumed, and several further key aspects were introduced
\cite{Sjostrand:1987su}. In its basic ideology this formalism has remained,
even if the details have been improved and extended many times over
the years.

This evolution will be described in the following, and in the process an
overview will be given of all the components of the current framework.
While \textsc{Pythia}-centered, external sources of inspiration
(in a positive or negative sense) will be mentioned, with emphasis
on the early days, when the basic ideas were formulated. Much more
information can be obtained from the companion articles of this book,
about other models and generators, and about all the experimental
studies that have been undertaken over the years. Notably, no experimental
plots are shown, since relevant ones are already reproduced elsewhere,
see \cite{CONTRIB:EXP-UE,CONTRIB:EXP-DPS-HF,CONTRIB:EXP-DPS-jets,%
CONTRIB:EXP-DPS-VB,CONTRIB:EXP-EventShape,CONTRIB:EXP-HM-PHENOMENOLOGY,%
CONTRIB:LOWX+DIFFRACTION,CONTRIB:EXP-MB-LHC},
often compared with \textsc{Pythia} and other generators.

\section{Early data and models}
\label{tssec:early}

In the eighties, the S$p\overline{p}$S was providing new data on
hadronic collisions, at an order of magnitude higher CM energy than
previously available, from 200 to 900~GeV. It came to change our
understanding of hadronic collisions. Some observations are of
special interest for the following.
\begin{Itemize}
\item The width $\sigma(n_{\mathrm{ch}})$ of the charged multiplicity
distribution is increasing with energy such that
$\sigma(n_{\mathrm{ch}}) / \langle n_{\mathrm{ch}} \rangle$ stays roughly
constant \cite{Alner:1984is,Ansorge:1988kn}, ``KNO scaling''
\cite{Koba:1972ng}, actually even slowly getting broader.
A close-to Poissonian process, in longitudinal phase space or in the
fragmentation of a single straight string, instead would predict a
$1 / \sqrt{\langle n_{\mathrm{ch}} \rangle}$ narrowing.
\item Multiplicity fluctuations show long-range ``forward--backward''
correlations \cite{Ansorge:1988fg}, defined by
\begin{equation}
b_{\mathrm{FB}}(\Delta \eta) = \frac{ \langle n_{\mathrm{F}}
n_{\mathrm{B}} \rangle - \langle n_{\mathrm{F}} \rangle^2}%
{ \langle n^2_{\mathrm{F}} \rangle - \langle n_{\mathrm{F}} \rangle^2} ~,
\end{equation}
where $n_{\mathrm{F}}$ and $n_{\mathrm{B}}$ is the (charged) multiplicity
in two symmetrically located unit-width pseudorapidity bins, separated
by a central variable-width $\Delta \eta$ gap. Again this is not
expected in Poissonian processes.
\item The average transverse momentum $\langle p_{\perp} \rangle$
increases with increasing charged multiplicity
\cite{Arnison:1982ed,Albajar:1989an}. This is opposite to the behaviour
at lower energies, where energy-momentum conservation effects dominate,
with a crossover at the highest ISR energies \cite{Breakstone:1983up}.
\item A non-negligible fraction of the total cross section is associated
with minijet production \cite{Ceradini:1985vm,Albajar:1988tt},
increasing from $\sim$5\% at 200~GeV to $\sim$15\% at 900~GeV.
Here UA1 defined a minijet as a region
$\Delta R = \sqrt{ (\Delta \eta)^2 + (\Delta \varphi)^2 } \leq 1$
with $\sum E_{\perp} > 5$~GeV.
\item The increase of the total $p\overline{p}$ cross section
$\sigma_{\mathrm{tot}}(s)$ rather well matches that of the minijet one
$\sigma_{\mathrm{minijet}}(s)$, \textit{i.e.}\
$\sigma_{\mathrm{tot}}(s) - \sigma_{\mathrm{minijet}}(s)$ is almost
constant \cite{Ceradini:1985vm,Albajar:1988tt}.
\item Events with a minijet have a rather flat
$\langle p_{\perp} \rangle (n_{\mathrm{ch}})$, while ones without show
a strong rise, starting from a lower level \cite{Ceradini:1985vm}.
\item The fraction of events having several minijets is non-negligible.
(Rates up to 5 are quoted from workshop presentations in
Ref.~\cite{Sjostrand:1987su}, but apparently never published.)
\item Events containing a hard jet also have an above-average level
of particle production well away from the jet core \cite{Arnison:1983gw},
the ``pedestal effect''. Of note is that the pedestal increases rapidly
up to $E_{\perp\mathrm{jet}} \sim 10$~GeV, and then flattens out, even
dropping slightly \cite{Albajar:1988tt}.
\item Also the jet profiles are affected by this extra source of activity.
\item By contrast, there were no early studies on double parton scattering
(DPS) at the S$p\overline{p}$S. The first observation instead came from
AFS at ISR \cite{Akesson:1986iv}, in a study of pairwise balancing jets
in four-jet events, but it did not convince everybody.
\end{Itemize}

On the theoretical side, the basic idea of MPI existed
\cite{Landshoff:1978fq,Goebel:1979mi,Paver:1982yp,Paver:1983hi,%
Humpert:1983pw,Humpert:1984ay,Paver:1984ux}, see also
\cite{CONTRIB:TH-DPS-HISTORICALOVERVIEW}. These first studies almost
exclusively considered DPS, without a vision of an arbitrary number of
scatterings. Studies often only included scattering of valence quarks,
since the large-$x$ region was needed to access ``large'' jet $p_{\perp}$
scales. Therefore DPS/MPI was only expected to correspond to a tiny
fraction of the total cross section. If needed, a $p_{\perp\mathrm{min}}$
cutoff would be introduced at a sufficiently high value to make it so.

For soft physics, the Pomeron language was predominant, notably in
its Dual Topological Unitarization (DTU) formulation, both to describe
total cross sections and event topologies
\cite{Gribov:1968fc,Abramovsky:1973fm,Veneziano:1974ag,Chew:1977yk,%
Capella:1978ig,Minakata:1979du,CohenTannoudji:1979mk,Fialkowski:1981nc,%
Aurenche:1982ne,Capella:1982yc,Kaidalov:1982xg,Kaidalov:1982xe}.
In it a cut Pomeron corresponds to two multiperipheral chains, or
strings in Lund language, stretched directly between the two beam
remnants after the collision. In most of the earlier phenomenological
studies only one cut Pomeron was used, but extensions to multiple
Pomerons were introduced for S$p\overline{p}$S applications.
Then the number of cut Pomerons can vary freely, \textit{e.g.}\ according
to a Poissonian. Uncut Pomerons, \textit{i.e.} virtual corrections,
ensure unitarity. This approach was quite successful in describing
aspects of the data such as the charged multiplicity distribution and
forward--backward correlations.

In contrast to the unitarization approach, the good match between the
rise of $\sigma_{\mathrm{tot}}(s)$ and $\sigma_{\mathrm{minijet}}(s)$
led to speculations that $\sigma_{\mathrm{tot}}(s)$ (or at least its
inelastic component) could be written as an incoherent sum
$\sigma_{\mathrm{tot}}(s) = \sigma_{\mathrm{soft}} + \sigma_{\mathrm{minijet}}(s)$
\cite{Gaisser:1984pg,Pancheri:1985ix,Martin:1985ge}.

At the time, there appears to have been little ``middle ground'' between
the hard MPI, the soft multi-Pomeron and the UA1-minijet ways of
approaching physics.

On the generator side, \textsc{IsaJet} \cite{Paige:1985tj} was state of
the art. It described one hard interaction with its showers, and then
added an underlying event (UE) based on the Pomeron approach. The UE was
intended to reproduce minimum-bias (MB) event properties at the
hard-interaction-reduced collision energy. Since it was based on
independent fragmentation the two components could be easily decoupled.
Other generators for hard interactions \cite{Odorico:1983yf,Field:1985tg}
had more primitive UE descriptions, and the one generator for MB
\cite{Aurenche:1983ek} did not include hard interactions. In addition
(unpublished) longitudinal phase-space models tuned to inclusive data
were used within the experimental collaborations, ultimately refined
into the UA5 generator \cite{Alner:1986is}.

\section{The first PYTHIA model}
\label{tssec:firstmodel}

Against this backdrop, the key new idea of the first \textsc{Pythia} model
\cite{Sjostrand:1985vv} was to reinterpret the multi-Pomeron picture in
terms of multiple perturbative QCD interactions. Thus there would no
longer be the need for separate descriptions of MB and UE physics.
A hard-process event would just be the high-$p_{\perp}$ tail of the MB
class, and a soft-process event just one where the hardest jet was too
soft to detect as such. MPIs come out as an unavoidable consequence,
not only as a tiny tail of hard DPS events, but as representing the
bulk of the inelastic nondiffractive cross section $\sigma_{\mathrm{nd}}$
at higher energies.

By contrast, no importance could be attached to the 5~GeV UA1 minijet
cutoff scale or to the seemingly simple relationship between
$\sigma_{\mathrm{tot}}(s)$ and $\sigma_{\mathrm{minijet}}(s)$ that it led to.
On the contrary, MPIs had to extend much lower in $p_{\perp}$ in order to
give enough varying activity to describe \textit{e.g.}\ the approximate
KNO scaling. Here jet universality was assumed, \textit{i.e.}\ that the
underlying fragmentation mechanism was the same string as described
$e^+e^-$ data so well, only applied to a more complicated partonic state.

\begin{figure}[t]
\centerline{\includegraphics[width=110mm]{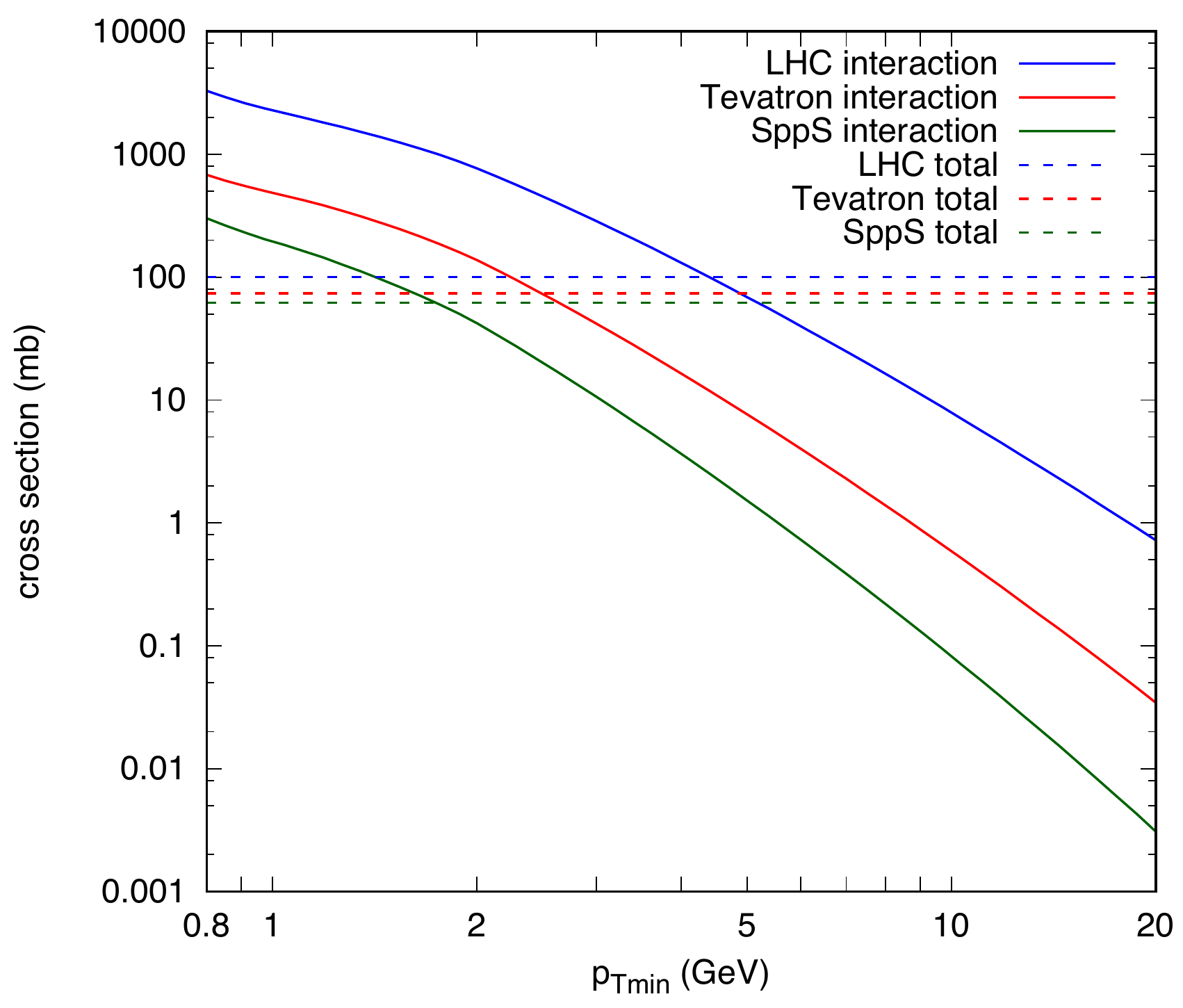}}
\caption{The integrated interaction cross section
$\sigma_{\mathrm{int}}(p_{\perp\mathrm{min}})$ for the S$p\overline{p}$S at
630~GeV, Tevatron at 1.96~TeV and LHC at 13~TeV. For comparison
the total cross section $\sigma_{\mathrm{tot}}$ at the respective energy
is indicated by a horizontal line, with the nondiffractive part
$\sigma_{\mathrm{nd}}$ at order 60\% of this. Results have been obtained
with the \textsc{Pythia}~8.223 default values, including the
NNPDF2.3 QCD+QED LO PDF set with $\alpha_{\mathrm{s}}(M_Z) = 0.130$
\cite{Ball:2013hta}. }
\label{tsfig:sigmaint}
\end{figure}

In its technical implementation, the starting point of the model is the
differential perturbative QCD $2 \to 2$ cross section
\begin{equation}
\frac{\mathrm{d}\sigma}{\mathrm{d} p_{\perp}^2} =
\sum_{i,j,k} \iiint f_i(x_1, Q^2) \, f_j(x_2, Q^2) \,
\frac{\mathrm{d}\hat{\sigma}_{ij}^k}{\mathrm{d}\hat{t}} \,
\delta \left( p_{\perp}^2 - \frac{\hat{t}\hat{u}}{\hat{s}} \right)
\, \mathrm{d}x_1 \, \mathrm{d}x_2 \, \mathrm{d}\hat{t} ~,
\label{tseq:dsigma}
\end{equation}
with $Q^2 = p_{\perp}^2$ as factorization and renormalization scale.
The corresponding integrated cross section depends on the chosen
$p_{\perp\mathrm{min}}$ scale:
\begin{equation}
\sigma_{\mathrm{int}}(p_{\perp\mathrm{min}}) =
\int_{p_{\perp\mathrm{min}}^2}^{s/4} \frac{\mathrm{d}\sigma}{\mathrm{d}p_{\perp}^2}
\, \mathrm{d}p_{\perp}^2~,
\label{tseq:sigint}
\end{equation}
see Fig.~\ref{tsfig:sigmaint}.

Diffractive events presumably give a small fraction of the perturbative
jet activity, and elastic none, so the simple model sets out to
describe only inelastic nondiffractive events, with an approximately
known $\sigma_{\mathrm{nd}}$. It is thus concluded that the average such
event ought to contain
\begin{equation}
\langle n_{\mathrm{MPI}}(p_{\perp\mathrm{min}}) \rangle =
\frac{\sigma_{\mathrm{int}}(p_{\perp\mathrm{min}})}{\sigma_{\mathrm{nd}}}
\label{tseq:nMPIavg}
\end{equation}
hard interactions. An average above unity corresponds to more
than one such subcollision per event, which is allowed by the
multiparton structure of the incoming hadrons. If the interactions
were to occur independently of each other,
$n_{\mathrm{MPI}}(p_{\perp\mathrm{min}})$ would be distributed according to
a Poissonian. But such an approach would be flawed, \textit{e.g.}\ sometimes
using up more energy for collisions than is available.

The solution to this problem was inspired by the
parton-shower paradigm. The generation of consecutive MPIs
is formulated as an evolution downwards in $p_{\perp}$,
resulting in a sequence of $n$ interactions with $\sqrt{s}/ 2 >
p_{\perp 1} > p_{\perp 2} > \cdots > p_{\perp n} > p_{\perp\mathrm{min}}$.
The probability distribution for $p_{\perp 1}$ becomes
\begin{equation}
\frac{\mathrm{d}\mathcal{P}}{\mathrm{d}p_{\perp 1}} =
\frac{1}{\sigma_{\mathrm{nd}}} \frac{\mathrm{d}\sigma}{\mathrm{d}p_{\perp 1}} \,
\exp \left( - \int_{p_{\perp 1}}^{\sqrt{s}/2} \frac{1}{\sigma_{\mathrm{nd}}}
\frac{\mathrm{d}\sigma}{\mathrm{d}p'_{\perp}} \, \mathrm{d}p'_{\perp} \right) ~.
\label{tseq:phardest}
\end{equation}
Here the naive probability is corrected by an exponential factor
expressing that there must not be any interaction in the range
between $\sqrt{s}/2$ and $p_{\perp 1}$ for $p_{\perp 1}$ to be the hardest
interaction. The procedure can be iterated, to give
\begin{equation}
\frac{\mathrm{d}\mathcal{P}}{\mathrm{d}p_{\perp i}} =
\frac{1}{\sigma_{\mathrm{nd}}} \frac{\mathrm{d}\sigma}{\mathrm{d}p_{\perp i}} \,
\exp \left( - \int_{p_{\perp i}}^{p_{\perp i - 1}} \frac{1}{\sigma_{\mathrm{nd}}}
\frac{\mathrm{d}\sigma}{\mathrm{d}p'_{\perp}} \, \mathrm{d}p'_{\perp} \right) ~.
\label{tseq:pnext}
\end{equation}
The exponential factors resemble Sudakov form factors of parton showers
\cite{Sudakov:1954sw,Buckley:2011ms}, or uncut Pomerons for that matter,
and fills the same function of ensuring probabilities bounded by unity.
Summing up the probability for a scattering at a given $p_{\perp}$ scale
to happen at any step of the generation chain gives back
$(1 / \sigma_{\mathrm{nd}}) \, \mathrm{d}\sigma / \mathrm{d}p_{\perp}$,
and the number of interactions above any $p_{\perp}$ is a Poissonian with
an average of $\sigma_{\mathrm{int}}(p_{\perp}) / \sigma_{\mathrm{nd}}$,
as it should. The downwards evolution in $p_{\perp}$ is routinely handled
by using the veto algorithm \cite{Sjostrand:2006za}, like for showers.

The similarities with showers should not be overemphasized, however.
While the shower $p_{\perp}$ scale has some approximate relationship to
an evolution in time, this is not so for MPIs. Rather, when the two
Lorentz-contracted hadron ``pancakes'' collide, the MPIs can be viewed
as occurring simultaneously in different parts of the overlap region.
What is instead gained is a way to handle the parton distribution
functions (PDFs) of several partons in the same hadron, at the very
least to conserve overall energy and momentum. Specifically, it is for
the hardest MPI that conventional PDFs have been tuned and tested,
so we had better respect that. For subsequent MPIs no PDF data exist,
so some adjustments are acceptable. In this first implementation only
rescaled PDFs
\begin{equation}
f(x'_i, Q^2)~~\mathrm{with}~~x'_i = \frac{x_i}{1 - \sum_{j = 1}^{i - 1} x_j} < 1
\label{tseq:simplefrescale}
\end{equation}
are used for the $i$'th interaction. This rescaling suppresses the tail
towards events with many MPIs, so the $n_{\mathrm{MPI}}$ distribution
becomes narrower than Poissonian.

To complement the model, a number of further details of the simulation
had to be specified, often intended as temporary solution.
\begin{Itemize}
\item There is a finite probability that no MPIs at all are generated above
$p_{\perp\mathrm{min}}$. For this set of events, small but not negligible,
an infinitely soft gluon exchange is assumed, leading to two strings
stretched directly between the beam remnants.
\item Only the hardest interaction is allowed to be any combination of
incoming and outgoing flavours, weighted according to
Eq.~(\ref{tseq:dsigma}). For subsequent ones kinematics is chosen the
same way, with modified PDFs, but afterwards the process is always set up
to be of the $g g \to g g$ type. The reason is to avoid complicated
beam-remnant structures.
\item The colour flow of the hardest interaction is described by the
original \textsc{Pythia} algorithm \cite{Bengtsson:1982jr}. Two extreme
scenarios for the colour flow of the non-hardest MPIs were compared.
In the simplest one, each such gives rise to a double string stretched
between the two outgoing gluons of the MPI, disconnected from the rest
of the event.
\item By the choices above, where only the hardest interaction
affects the flavour and colour of the beam remnant, a limited number of
remnant types can be obtained. If a valence quark is kicked out, a diquark
is left behind. If a gluon, the leftover colour octet state of a proton
can be split into a quark and a diquark that attach to two separate strings.
If the two remnants then share the longitudinal momentum evenly, it
maximizes the particle production. This gives too few low-multiplicity
events, and also leaves less room for MPIs to build up a high-multiplicity
tail, assuming that the average is kept fixed. Therefore a probability
distribution is used wherein the quark often obtains much less momentum
than the diquark. Finally, if a sea (anti)quark is kicked out, the remnant
can be split into a single hadron plus a quark or diquark.
\item Only the hardest interaction is dressed up with showers,
whereas the subsequent ones are not. Again the reason is beam-remnant
issues, but one excuse is that most non-hardest MPIs appear at low
$p_{\perp}$ scales, where little further radiation should be allowed.
\end{Itemize}

The key free parameter of this framework is $p_{\perp\mathrm{min}}$.
The lower it is chosen, the higher the average number of MPIs,
cf.\ Eq.~(\ref{tseq:sigint}), and thus the higher the average charged
multiplicity $\langle n_{\mathrm{ch}} \rangle$. To agree with 540~GeV
UA5 data \cite{Alner:1984is} a value of $p_{\perp\mathrm{min}} = 1.6$~GeV
was required. The dependence of $\langle n_{\mathrm{ch}} \rangle$ on
$p_{\perp\mathrm{min}}$ is quite strong so, if everything else is kept
fixed, the $p_{\perp\mathrm{min}}$ uncertainty is of order $\pm 0.1$~GeV.
The most agreeable aspect, however, is that with $p_{\perp\mathrm{min}}$
tuned, the shape of the $n_{\mathrm{ch}}$ distribution is reasonably
well described. Without MPIs the shape is Poissonian-like, also when
a single hard interaction is allowed. With MPIs instead an approximate
KNO scaling behaviour is obtained, driven by the $n_{\mathrm{MPI}}$
distribution. (Which, even if also a Poissonian, obeys
$\langle n_{\mathrm{MPI}} \rangle \ll \langle n_{\mathrm{ch}} \rangle$,
meaning much larger relative fluctuations $\sigma / \langle n \rangle$.)
By the same mechanism also strong forward--backward correlations are
obtained, where before these were tiny. That is, the $n_{\mathrm{MPI}}$
is a kind of global quantum number of an event, that affects whether
particle production is high or low over the whole rapidity range.
With some damping, since not all strings are stretched equally far out
to the beam remnants.

In part this is nothing new; the number of cut Pomerons in soft models
fills a similar function for both $n_{\mathrm{MPI}}$ distributions and
forward--backward correlations. What is new is that an application of
perturbation theory, in combination with string fragmentation, can give
a reasonable description also of minimum-bias physics.
This unifies hard and soft physics at colliders, as being part of the
same framework. It also introduces a new cutoff scale in QCD, with a
value different from other scales, such as the proton mass.

It was clear from the onset that the model was incomplete in its
details, and the listed open questions for the model well matches the
problems that have later been studied.
\begin{Itemize}
\item What is the correct behaviour of
$\mathrm{d}\sigma/\mathrm{d} p_{\perp}^2$ at small $p_{\perp}$?
A sharp cutoff, below which cross sections vanish, is not plausible.
\item How to remove (or, if not, interpret) the class of events with
no MPIs, currently represented by a $p_{\perp} = 0$ interaction?
\item How to introduce an impact-parameter picture, giving more activity
for central collisions and less for peripheral? This is needed to give
an a bit wider $n_{\mathrm{ch}}$ distribution. Also, for UA1 jets
the MPI formalism as it stands at this stage only gives about a quarter
of the observed pedestal effect.
\item How to achieve a better description of multiparton PDFs, that also
consistently includes \textit{e.g.}\ flavour conservation and correlations?
\item Where does the baryon number go if several valence quarks are
kicked out from a proton?
\item How does the colour singlet nature of the incoming beams translate
into colour correlations between the different MPIs?
\item What is the structure and role of beam remnants?
\item By confinement and the uncertainty relation the incoming partons
must have some random nonperturbative transverse motion. How should such
``primordial $k_{\perp}$'' effects be included? These then have to be
compensated in the remnants, and furthermore the remnant parts may have
relative  $k_{\perp}$ values of their own.
\item How should parton-shower effects be combined consistently between
the systems? The flavour, colour and beam-remnant issues reappear here.
\item How important is ISR evolution wherein a parton branches into two
that participate in two separate interactions?
\item How important is rescattering, \textit{i.e.}\ when one parton can
scatter consecutively from two or more partons from the other hadron?
\item How do diffractive topologies contribute to the picture?
Typical experimental ``minimum bias'' triggers catch a fraction of
these events, which have different properties from the nondiffractive
ones. The low-multiplicity end of the $n_{\mathrm{ch}}$ distribution
was left unexplained in the studies, with the motivation that it is
dominated by diffraction.
\item How do the results scale with collision energy? With a fixed
$p_{\perp\mathrm{min}}$ scale it was possible to reproduce the
$\langle n_{\mathrm{ch}} \rangle$ evolution from fixed-target to 900~GeV,
and this was the basis for extrapolations.
\end{Itemize}

\section{Smooth damping and impact-parameter dependence}
\label{tssec:dampenimpact}

For the first published MPI article \cite{Sjostrand:1987su}, the
original framework was extended to address some of the most pressing
shortcomings above. (The older approach was also retained as a simpler
alternative. Unfortunately the new approach led to longer computer
generation times, which was a real issue at the time.)

The sharp cutoff $p_{\perp\mathrm{min}}$ is replaced by a smooth turnoff
at a scale $p_{\perp 0}$. To be specific, the cross section of
Eq.~(\ref{tseq:dsigma}) is multiplied by a damping factor
\begin{equation}
\left( \frac{\alpha_{\mathrm{s}}(p_{\perp 0}^2 + p_{\perp}^2)}%
{\alpha_{\mathrm{s}}(p_{\perp}^2)} \,
\frac{p_{\perp}^2}{p_{\perp 0}^2 + p_{\perp}^2} \right)^2 ~.
\label{tseq:ptdamp}
\end{equation}
Since the QCD $2 \to 2$ processes are dominated by $t$-channel gluon
exchange, which behaves like $1/\hat{t}^2 \sim 1 / p_{\perp}^4$, this
means that
\begin{equation}
\frac{\mathrm{d}\sigma}{\mathrm{d} p_{\perp}^2} \sim
\frac{\alpha_{\mathrm{s}}^2(p_{\perp}^2)}{p_{\perp}^4} \to
\frac{\alpha_{\mathrm{s}}^2(p_{\perp 0}^2 + p_{\perp}^2)}%
{(p_{\perp 0}^2 + p_{\perp}^2)^2}  ~,
\end{equation}
which is finite in the limit $p_{\perp} \to 0$. This behaviour can
be viewed as a consequence of colour screening: in the $p_{\perp} \to 0$
limit a hypothetical exchanged gluon would not resolve individual
partons but only (attempt to) couple to the vanishing net colour charge
of the hadron. Technically the damping factor is multiplying the
$\mathrm{d}\hat{\sigma} / \mathrm{d}\hat{t}$ expressions, but it
could equally well have been imposed (half each) on the PDFs instead,
since neither can be trusted for $p_{\perp} \to 0$.

In this modified framework all interactions are associated with a
$p_{\perp} > 0$ scale, and at least one interaction must occur when
two hadrons pass by for there to be an event at all. Thus we require
$\sigma_{\mathrm{int}}(0) > \sigma_{\mathrm{nd}}$, where the
$\sigma_{\mathrm{int}}$ integration, Eq.~(\ref{tseq:sigint}), now
includes the damping factor. A tune to $\langle n_{\mathrm{ch}} \rangle$
at 540~GeV gives $p_{\perp 0} \approx 2.0$~GeV, \textit{i.e.}\ of the same
order as the sharp $p_{\perp\mathrm{min}}$ cutoff. The two would have been
even closer, had not factorization and renormalization scales here been
multiplied by 0.075, as suggested at the time to obtain an approximate
NLO jet cross section \cite{Ellis:1985er}. Below S$p\overline{p}$S
energies a fixed $p_{\perp 0}$ gives too small a  $\sigma_{\mathrm{int}}(0)$,
so in this form the model is primarily intended for high-energy collider
physics.

The other main change was to introduce a dependence on the impact
parameter $b$ of the collision process. To do this, a spherically
symmetric matter distribution
$\rho(\mathbf{x}) \, \mathrm{d}^3x = \rho(r) \, \mathrm{d}^3x$
is assumed. In a collision process the overlap of the two hadrons
is then given by
\begin{eqnarray}
\widetilde{\mathcal{O}}(b) & = & \iint \mathrm{d}^3x \, \mathrm{d}t \,
\rho_{\mathrm{boosted}}\left( x - \frac{b}{2}, y, z - vt \right) \,
\rho_{\mathrm{boosted}}\left( x + \frac{b}{2}, y, z + vt \right)
\nonumber \\
& \propto &
\iint \mathrm{d}^3x \, \mathrm{d}t \, \rho ( x, y, z) \,
\rho ( x, y, z - \sqrt{b^2 + t^2} )  ~,
\label{tseq:convolute}
\end{eqnarray}
where the second line is obtained by suitable scale changes.

\begin{figure}[t]
\centerline{\includegraphics[width=110mm]{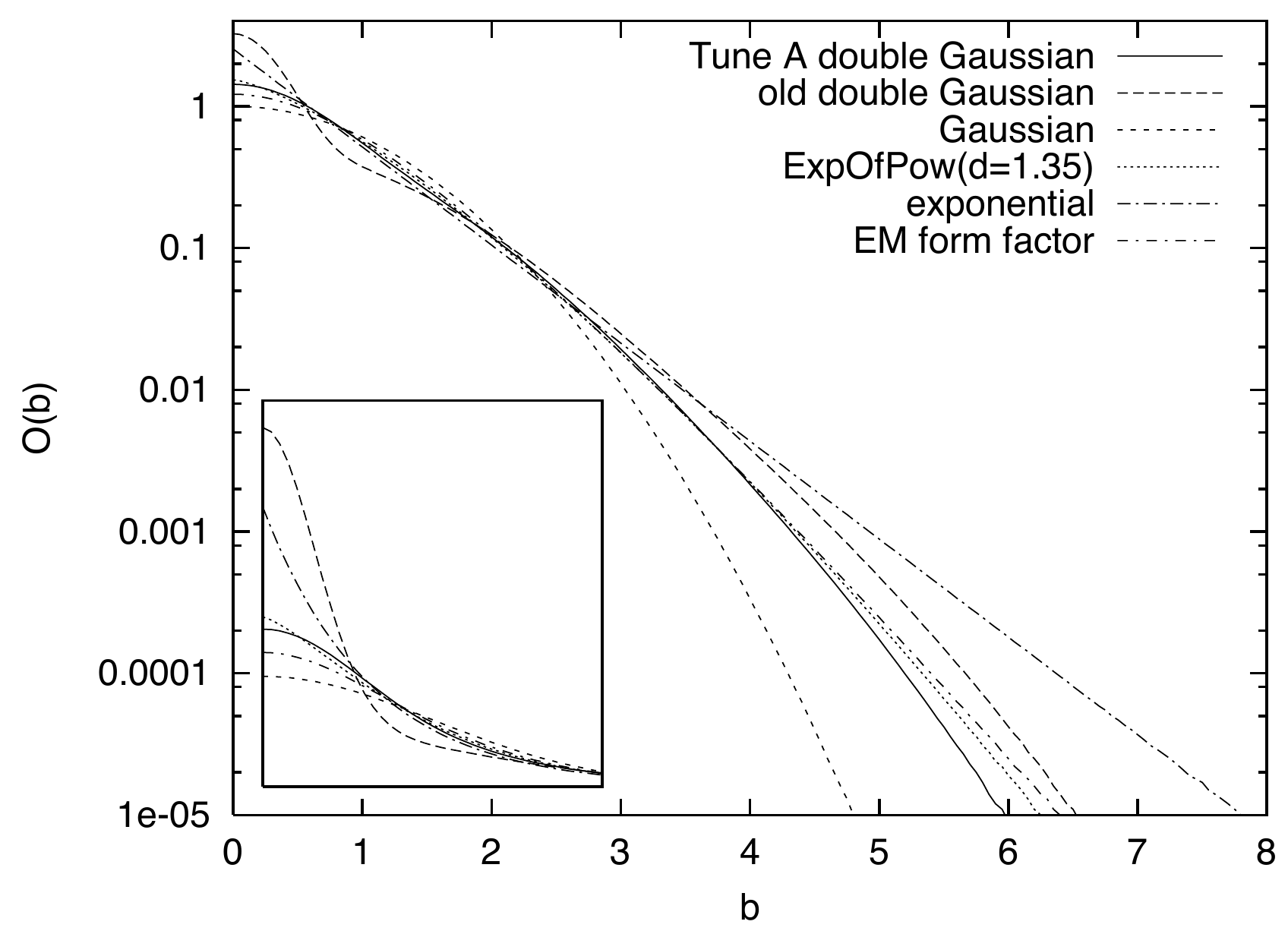}}
\caption{Examples of impact-parameter profiles $\widetilde{\mathcal{O}}(b)$,
some introduced only later. Somewhat arbitrarily the different
parametrizations have been normalized to the same area and average $b$,
\textit{i.e.}\ same $\int \widetilde{\mathcal{O}}(b) \, \mathrm{d}^2b$
and $\int b\widetilde{\mathcal{O}}(b) \, \mathrm{d}^2b$. Insert shows
the region $b < 2$ on a linear scale. From Ref.~\cite{Sjostrand:2004pf}.}
\label{tsfig:bprofile}
\end{figure}

A few different $\rho$ distributions were studied,
Fig.~\ref{tsfig:bprofile}. Using Gaussians is especially convenient,
since the convolution then becomes trivial. A simple Gaussian was the
starting point, but we found it did not give a good enough description
of the data. Instead the preferred choice was a double Gaussian
\begin{equation}
\rho(r) =
(1 - \beta) \, \frac{1}{a_1^3} \, \exp\left( - \frac{r^2}{a_1^2} \right)
+ \beta \, \frac{1}{a_2^3} \, \exp\left( - \frac{r^2}{a_2^2} \right) ~.
\end{equation}
This corresponds to a distribution with a small core region, of radius
$a_2$ and containing a fraction $\beta$ of the total hadronic matter,
embedded in a larger hadron of radius $a_1$.  The choice of a
not-so-smooth shape was largely inspired by the ``hot spot'' ideas
popular at the time \cite{Gribov:1984tu,Mueller:1985wy}. The starting
point is that, as a consequence of parton cascading, partons may tend to
cluster in a few small regions, typically associated with the three
valence quarks.

More convoluted ans\"atze could have been considered, but having two
free parameters, $\beta$ and $a_2 / a_1$, was sufficient to give
the necessary flexibility.

It is now assumed that the interaction rate, to first approximation,
is proportional to the overlap
\begin{equation}
\langle \widetilde{n}_{\mathrm{MPI}}(b) \rangle
= k \, \widetilde{\mathcal{O}}(b)~.
\end{equation}
For each given $b$ the number of interactions is assumed distributed
according to a Poissonian, at least before energy--momentum conservation
issues are considered. Zero interactions means that the hadrons pass
each other without interacting. The $\widetilde{n}_{\mathrm{MPI}}(b) \geq 1$
interaction probability therefore is
\begin{equation}
\mathcal{P}_{\mathrm{int}}(b) =
 1 - \exp \left( - \langle \widetilde{n}_{\mathrm{MPI}}(b) \rangle \right)
= 1 - \exp \left( - k \, \widetilde{\mathcal{O}}(b) \right) ~.
\label{tseq:bprob}
\end{equation}
We notice that $k \widetilde{\mathcal{O}}(b)$ is essentially the same as
the eikonal $\Omega(s,b) = 2\, \mathrm{Im}\chi(s,b)$ of optical models
\cite{Glauber:1959xx,Chou:1968bc,Bourrely:1984gi,LHeureux:1985qwr},
but split into one piece $\widetilde{\mathcal{O}}(b)$ that is purely
geometrical and one $k = k(s)$ that carries the information on the
parton--parton interaction cross section. Furthermore, this is (so far)
only a model for nondiffractive events, so does not attempt to relate
the eikonal to total or diffractive cross sections.

Simple algebra shows that the average number of interactions in events,
\textit{i.e.}\ hadronic passes with $n_{\mathrm{MPI}} \geq 1$, is given by
\begin{equation}
\langle n \rangle = \frac{\int k \, \widetilde{\mathcal{O}}(b) \, %
\mathrm{d}^2 b}{\int \mathcal{P}_{\mathrm{int}}(b) \, \mathrm{d}^2 b}
= \frac{\sigma_{\mathrm{int}}(0)}{\sigma_{\mathrm{nd}}} ~,
\end{equation}
which fixes the absolute value of $k$ (numerically).

For event generation, Eq.~(\ref{tseq:phardest}) generalizes to
\begin{equation}
\frac{\mathrm{d}\mathcal{P}}{\mathrm{d}^2 b \, \mathrm{d}p_{\perp 1}} =
\frac{\widetilde{\mathcal{O}}(b)}{\langle \widetilde{\mathcal{O}} \rangle}
\, \frac{1}{\sigma_{\mathrm{nd}}} \frac{\mathrm{d}\sigma}{\mathrm{d}p_{\perp 1}}
\, \exp \left( -
\frac{\widetilde{\mathcal{O}}(b)}{\langle \widetilde{\mathcal{O}} \rangle}
\int_{p_{\perp 1}}^{\sqrt{s}/2} \frac{1}{\sigma_{\mathrm{nd}}}
\frac{\mathrm{d}\sigma}{\mathrm{d}p'_{\perp}} \, \mathrm{d}p'_{\perp} \right) ~,
\label{tseq:phardestwithb}
\end{equation}
with the definition
\begin{equation}
\langle \widetilde{\mathcal{O}} \rangle =
\frac{\int \widetilde{\mathcal{O}}(b) \, \mathrm{d}^2 b}%
{\int \mathcal{P}_{\mathrm{int}}(b) \, \mathrm{d}^2 b} ~.
\end{equation}
Hence $\widetilde{\mathcal{O}}(b) / \langle \widetilde{\mathcal{O}} \rangle$
represents the enhancement at small $b$ and depletion at large $b$.
The simultaneous selection of $p_{\perp 1}$ and $b$ is somewhat more
tricky. In practice different schemes are used, depending on context.
\begin{Itemize}
\item For minimum-bias events Eq.~(\ref{tseq:phardestwithb}) can be
integrated oven $p_{\perp 1}$ to give $\mathcal{P}_{\mathrm{int}}(b)$ of
Eq.~(\ref{tseq:bprob}). To pick such a $b$ it is useful to note that
$\mathcal{P}_{\mathrm{int}}(b) < \min \left( 1, %
k \, \widetilde{\mathcal{O}}(b) \right)$ and split the $b$ range
accordingly. Once $b$ is fixed the selection of $p_{\perp 1}$ can be done
as for Eq.~(\ref{tseq:phardest}), only with an extra fix
$\widetilde{\mathcal{O}}(b) / \langle \widetilde{\mathcal{O}} \rangle$,
both in the prefactor and in the exponential.
If $p_{\perp 1} = 0$ is reached in the downwards evolution without an
interaction having been found, which happens with probability
$\exp \left( - k \, \widetilde{\mathcal{O}}(b) \right)$, then the
generation is restarted at the maximum scale $\sqrt{s} / 2$.
\item For a very hard process the exponential of
Eq.~(\ref{tseq:phardestwithb}) is very close to unity and can be dropped.
Then the selection of $b$ and $p_{\perp 1}$ decouples and can be done
separately. Here $\mathrm{d}\sigma / \mathrm{d}p_{\perp 1}$ can represent
any hard process, not only QCD jets, and $p_{\perp 1}$ any set of relevant
kinematic variables.
\item For a medium-hard process one can begin as in the hard case,
and then use the exponential as an acceptance probability. If the
hard-process kinematics is considered fixed then only a new $b$ value
is chosen in case of rejection. Note that it is always the QCD cross
section that enters in the exponential. (Or, to be proper, the sum of
all possible reactions, but that is completely dominated by QCD.)
For non-QCD processes the $p_{\perp 1}$ scale in the exponential has to
be associated with some suitable hardness scale, like the mass for the
production of a resonance.
\end{Itemize}

Once the hardest interaction is chosen, the generation of subsequent
ones proceeds by a logical extension of Eq.~(\ref{tseq:pnext}) to
\begin{equation}
\frac{\mathrm{d}\mathcal{P}}{\mathrm{d}p_{\perp i}} =
\frac{\widetilde{\mathcal{O}}(b)}{\langle \widetilde{\mathcal{O}} \rangle} \,
\frac{1}{\sigma_{\mathrm{nd}}} \frac{\mathrm{d}\sigma}{\mathrm{d}p_{\perp i}} \,
\exp \left( -
\frac{\widetilde{\mathcal{O}}(b)}{\langle \widetilde{\mathcal{O}} \rangle} \,
\int_{p_{\perp i}}^{p_{\perp i - 1}} \frac{1}{\sigma_{\mathrm{nd}}}
\frac{\mathrm{d}\sigma}{\mathrm{d}p'_{\perp}} \, \mathrm{d}p'_{\perp} \right) ~.
\label{tseq:pnextwithb}
\end{equation}

There is one subtlety to note about ordering, however. QCD interactions
have to be ordered in $p_{\perp}$ for the formalism to reproduce the
correct inclusive cross section. This applies for the MB generation,
which gives an arbitrary $p_{\perp 1}$, and also in a sample of hard
jets above some large $p_{\perp\mathrm{min}}$ scale. But it does not hold
for non-QCD hard processes. For $Z^0$ production, say, which is not part
of the normal MPI machinery, the second MPI (counting the $Z^0$ as the
first) can go all the way up to the kinematic limit in $p_{\perp}$
without involving any double-counting, with $p_{\perp}$-ordering only
kicking in for the third MPI.

With MPIs stretching down to $p_{\perp} = 0$, the need arises to evaluate
PDFs below their lower limit $Q_0$ scale, typically 1 -- 2~GeV. To first
approximation this is done by freezing them below $Q_0$, but some
attempts were made to enhance the relative importance of valence quarks
for $Q \to 0$, since this is what one should expect to happen.

\begin{figure}[t]
\centerline{\includegraphics[width=110mm]{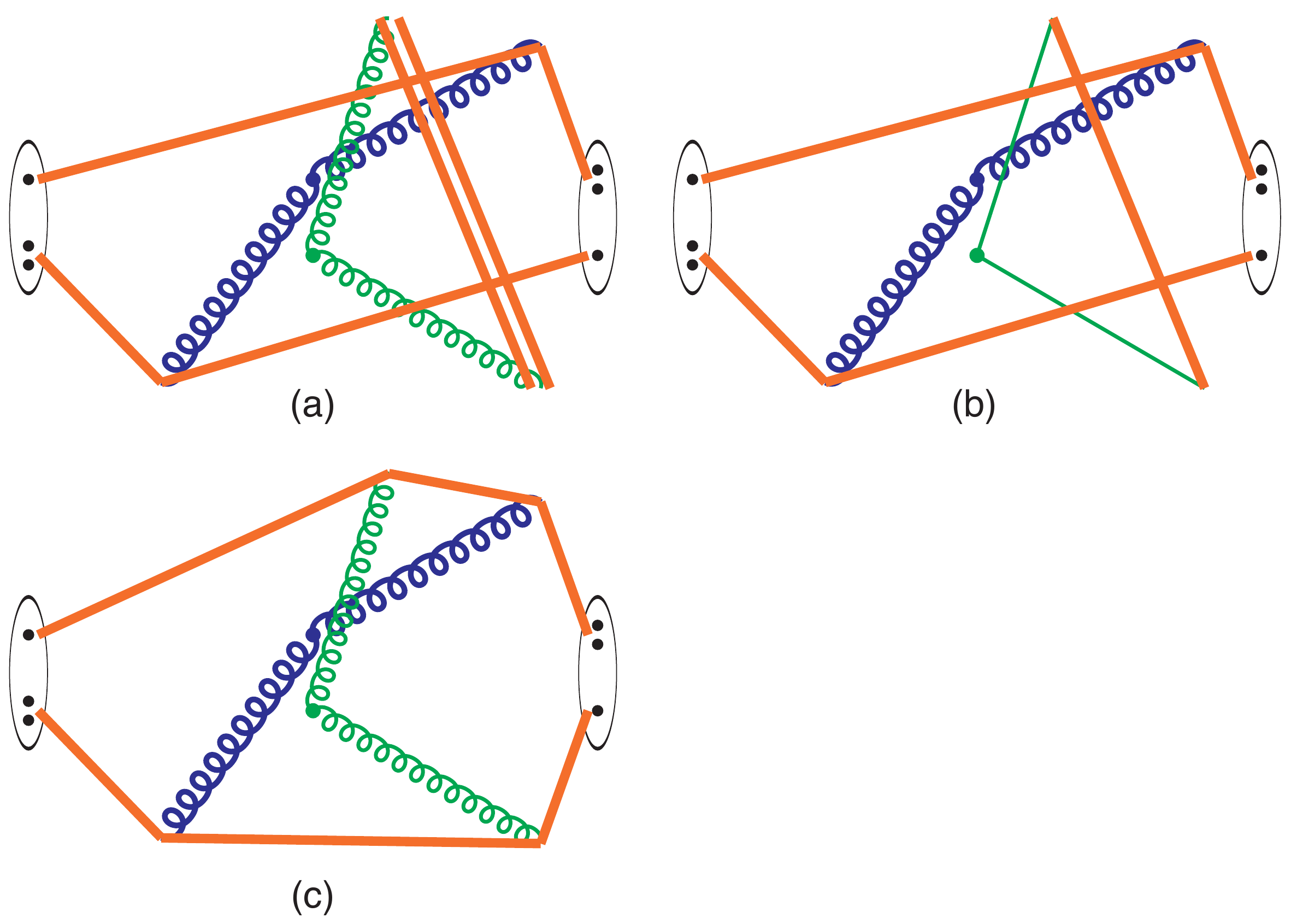}}
\caption{Colour drawing possibilities in the final state for the simple
model. Thick blue gluons denote outgoing partons from the primary
interaction, thin green gluons or quark lines the partons of a further MPI,
black ovals the beam remnants with valence quarks, and orange thick lines
the colour strings stretched between the partons. While the primary
interaction and its connection to the beam remnants is handled according
to the colour flow of the matrix elements, in the $N_C \to \infty$ limit
\cite{tHooft:1973alw}, the further MPIs give a mix of behaviours (a), (b)
and (c), as described in the text. Note that the figure is not to scale;
\textit{e.g.}\ that the strings have a transverse width of hadronic size.}
\label{tsfig:colourdraw}
\end{figure}

As before, colour drawing for all MPIs except the hardest one is handled
in a primitive manner. Given that the kinematics of an interaction has
been chosen with the full cross section, the final state is picked
among three possibilities, Fig.~\ref{tsfig:colourdraw}, by default with
equal probability.
\begin{Enumerate}
\item[(a)] Assume the collision to have produced a $gg$ pair and stretch
two string pieces directly between them, giving a closed gluon loop.
\item[(b)] Assume the collision to have produced a $q\overline{q}$ pair and
stretch a string directly between them.
\item[(c)] Assume the collision to have produced a $gg$ pair, but insert
them separately on an already existing string in such a way so as to
minimize the increase of string length $\lambda$ \cite{Andersson:1985qr}.
Here
\begin{equation}
\lambda \approx
\sum_{i = 0}^{n} \ln \left( 1 + \frac{ m_{i,i+1}^2}{m_0^2} \right) ~,
~~~~ m_{i,i+1}^2 = (\epsilon_i p_i + \epsilon_{i+1} p_{i+1})^2 ~,
\label{tseq:lambda}
\end{equation}
for a string $q_0 g_1 g_2 \cdots g_n \overline{q}_{n+1}$, where
$\epsilon_q = 1$ but $\epsilon_g = 1/2$ because a gluon momentum is
shared between two string pieces it is connected to.
\end{Enumerate}
Neither of these three follow naturally from any colour flow rules,
such as $t$-channel gluon exchange. Rather the first two represent the
simplest way to decouple different interaction systems from each other,
not having to trace colours back through the beam remnants.
If MPIs are such separated systems, and thus on the average each gives
the same $\langle p_{\perp} \rangle$, then an essentially flat
$\langle p_{\perp} \rangle (n_{\mathrm{ch}})$ would be expected, since the
study of the $n_{\mathrm{ch}}$ distribution tells us that higher
$n_{\mathrm{ch}}$ values is a consequence of more MPIs rather than of
harder jets. To obtain a rising
$\langle p_{\perp} \rangle (n_{\mathrm{ch}})$ it is therefore
essential to have a mechanism to connect the different MPI subsystems
in colour, not only at random but specifically so as to reduce the total
string length of the event, more and more the more MPIs there are.
Each further MPI on the average then contributes less $n_{\mathrm{ch}}$
than the previous, while still the same (semi)hard $p_{\perp}$ kick is to
be shared between the hadrons, thus inducing the rising trend.
This is precisely what the third and last component is intended to do.
It is the first large-scale application of colour reconnection (CR) ideas,
previous applications having been for more specific channels such as
$B \to J/\psi$ decays \cite{Fritzsch:1977ay,Ali:1978kn,Fritzsch:1979zt}.

A very simple model for diffraction was also added, wherein the
diffractive mass $M$ is selected according to $\mathrm{d}M^2 / M^2$
and is represented by a single string stretched between a diquark in
the forward direction and a quark in the backward one.

With these changes to the original model it is possible to obtain a
quite reasonable description of essentially all the key experimental
data outlined in Sec.~\ref{tssec:early}. Above all, the model offered
physics explanations for the behaviour observed in data.
\begin{Itemize}
\item For the charged multiplicity distribution, improvements in the
high-multiplicity tail originate from the introduction
of an impact-parameter picture, whereas the addition of diffraction
improves the low-multiplicity one. To describe the energy dependence,
where $\sigma(n_{\mathrm{ch}}) / \langle n_{\mathrm{ch}} \rangle$ is slightly
increasing with energy, the impact-parameter dependence is crucial,
since the $\sigma(n_{\mathrm{MPI}}) / \langle n_{\mathrm{MPI}} \rangle$
then does not fall, which it otherwise would when
$\langle n_{\mathrm{MPI}} \rangle$ increases with energy.
Also forward-backward correlations now are even stronger, reflecting
the broader $n_{\mathrm{MPI}}$ distribution, and actually even somewhat
above data. A number of other minimum-bias distributions look fine, like
the $\mathrm{d} n_{\mathrm{ch}} / \mathrm{d} \eta$ spectrum, inclusively
as well as split into multiplicity bins, except for the lowest one,
which is dominated by diffraction.
\item The $\langle p_{\perp} \rangle (n_{\mathrm{ch}})$ distribution is
well described, both inclusively and split into samples with our without
minijets. As already mentioned, the new CR mechanism here plays a key role
to get the correct rising trend in the inclusive case, and to counteract
the drop otherwise expected in the jet sample. Not only the slope but
also the absolute value of $\langle p_{\perp} \rangle$ is well reproduced,
without any need to modify the fragmentation $p_{\perp}$ width tuned to
$e^+e^-$ data. This is one of the key observations that lend credence
to the jet universality concept.
\item The UA1 minijet studies are rather well reproduced. Notably the
default double Gaussian is needed to obtain the observed fraction with
several minijets. The simpler alternatives with a single Gaussian, no
$b$ dependence, or no MPIs at all fared increasingly worse, even with
$\alpha_{\mathrm{s}}$ tuned to give the same average number of minijets.
\item The pedestal effect, \textit{i.e.}\ how the underlying event activity
first rises with the trigger jet/cluster $E_{\perp}$, is well described,
and explained. The rise is is caused by a shift in the composition of
events, from one dominated by fairly peripheral collisions to one
strongly biased towards central ones. In the model there is a limit
for how far this biasing can go: the exponential in
Eq.~(\ref{tseq:phardestwithb})) can be neglected once
$p_{\perp 1} \simeq E_{\perp}$ is so large that
$\sigma_{\mathrm{int}}(p_{\perp 1}) \ll \sigma_{\mathrm{nd}}$.
This happens at around 10~GeV, explaining the origin of that scale.
The probability distribution is then given by
$\widetilde{\mathcal{O}}(b) \, \mathrm{d}^2 b$ independently of the
$p_{\perp 1}$ value. The double Gaussian is required to obtain the
correct pedestal height, whereas a single Gaussian undershoots. A slight
drop of the pedestal height for $E_{\perp} > 25$~GeV can be attributed
to a shift from mainly $gg$ interactions to mainly $q\overline{q}$ ones.
\end{Itemize}

In summary, most if not all of MB and UE physics at collider energies
is explained and reasonably well described once the basic MPI framework
has been complemented by a smooth turnoff of the cross section for
$p_{\perp} \to 0$, a requirement to have at least one MPI to get an event,
an impact-parameter dependence, and a colour reconnection mechanism.

\section{Interlude}
\label{tssec:interlude}

While the S$p\overline{p}$S had paved the way for a new view on hadronic
collisions, the Tevatron rather contributed to cement this picture.
KNO distributions kept on getting broader \cite{Alexopoulos:1998bi},
forward-backward correlations got stronger \cite{Alexopoulos:1995ft},
and $\langle p_{\perp} \rangle (n_{\mathrm{ch}})$ showed the same rising
trend \cite{Alexopoulos:1994ag,Acosta:2001rm}, to give some examples.
The Tevatron emphasis was on hard physics, however, and it took many
years to go beyond the S$p\overline{p}$S MB and UE studies.
Notable is the CDF study on the production of $\gamma$ + 3 jets
\cite{Abe:1997xk}, which came to be the first generally accepted proof
of the existence of DPS. Studies of the pedestal effect eventually also
became quite sophisticated \cite{Affolder:2001xt,Field:2002vt,
Acosta:2004wqa,Kar:2008zza}, providing differential information on
activity in towards, away and transverse regions in azimuth relative to
the trigger, including a $Z^0$ trigger. All of these observations were
in qualitative agreement with \textsc{Pythia} predictions.
An improved quantitative agreement was obtained in a succession of tunes
\cite{Field:2005sa,Field:2006iy}, see also
\cite{CONTRIB:EXP-UE}, like the much-used Tune A.

Even if agreement may not have been perfect, there was no obvious
pattern of disagreement between S$p\overline{p}$S/Tevatron data and
the \textsc{Pythia} model. Therefore it could routinely be used for
experimental studies and for extrapolations to LHC and SSC energies.
But it also meant that further MPI development was slow in the period
1988 -- 2003, with only some relevant points, as follows.

More up-to-date formulae for total, elastic and diffractive cross
sections were implemented \cite{Schuler:1993wr}, starting from the
$\sigma_{\mathrm{tot}}(s)$ parametrizations of Donnachie and Landshoff
(DL) \cite{Donnachie:1992ny}. They are based on an effective Pomeron
description, with parameters adjusted to describe existing data and
also give a reasonable extrapolation to high energies. They worked well
through the Tevatron era, but overestimated the diffractive rate for
LHC and have since been modified.

The $p_{\perp 0}$ parameter went through several iterations, as new
PDF sets appeared on the market and became defaults in \textsc{Pythia}.
Notably HERA data showed that there is a non-negligible rise in the
small-$x$ region, even for small $Q^2$ scales, whereas pre-HERA PDFs
had tended to enforce a flat $xf(x, Q_0^2)$ at small $x$. This implies
that the all-$p_{\perp}$ integrated QCD cross section rises much faster
with $s$ than assumed before, and thereby generates a faster rising
$\langle n_{\mathrm{ch}} \rangle (s)$. The need for an $s$ dependence of
$p_{\perp 0}$, which previously had been marginal, now became obvious.
Initially a logarithmic $s$ dependence was used. Later on a power-like
form was introduced, such as
\begin{equation}
p_{\perp 0}(s) = (2.1~\mathrm{GeV}) \left( \frac{s}{1~\mathrm{TeV}^2}
\right)^{\epsilon} \,
\end{equation}
with $\epsilon = 0.08$, inspired by the DL ansatz
$\sigma_{\mathrm{tot}}(s) \simeq s^{\epsilon}$, which also qualitatively
matches well with a HERA $xf(x, Q_0^2) \simeq x^{-\epsilon}$ behaviour.

In an attempt to understand the behaviour of $p_{\perp 0}(s)$, a simple
toy study was performed \cite{Dischler:2000pk}. As we already argued,
the origin of a $p_{\perp 0}$ damping scale in the first place is that
the proton is a colour singlet, which means that individual parton
colour charges are screened. A very naive estimate is that the screening
distance should be the inverse of the proton size,
$p_{\perp 0} \approx \hbar / r_p \approx 0.3$~fm. But this assumes
that the proton only consists of very few partons, such that the typical
distance between two partons is $r_p$. In reality we expect the evolution
of PDFs, especially at small $x$, to lead to a much higher density.
Therefore the typical colour screening distance --- how far away you
need to go to find partons with opposing colour charges --- to be much
smaller than $r_p$. In order to test this, we built a model for the
transverse structure of the proton as follows. Start out from a picture
with three valence quarks and two ``valence gluons'' that share the full
momentum of the proton at a scale $Q_0 \approx 0.5$~GeV, based on the
GRV PDF approach \cite{Gluck:1994uf}, distributed across a transverse
proton disc, and with net vanishing colour. They are then evolved upwards
in $Q^2$, to create ISR cascades. A technical problem is that the
$x \to 0$ singularity would generate infinitely many partons. Therefore
branchings are only allowed if both daughters have an
$x > x_{\mathrm{min}} \simeq p_{\perp 0}/\sqrt{s}$. Colours are preserved
in branchings, and daughters can drift a random amount in transverse
space of order $\hbar / Q$ if produced at a scale $Q$. A damping factor
can then be defined by
\begin{equation}
\frac{ \left| \sum_k q_k \, e^{i\mathbf{r}_k \mathbf{p}_{\perp}} \right|^2 }%
{\sum_k |q_k|^2}~,
\end{equation}
where $\mathbf{p}_{\perp}$ represents a gluon plane wave probing the
proton, consisting of partons with colour charge $q_k$ located at
$\mathbf{r}_k$. This approach indeed gives results consistent with a
damping at scales around 2~GeV, varying with $s$ about as outlined
above, but it contains too many uncertainties to be used for any
absolute predictions.

The MPI framework was extended to $\gamma p$ \cite{Schuler:1993td} and
$\gamma \gamma$ \cite{Schuler:1996en} collisions. It there was applied
to the Vector Meson Dominance (VMD) part of the photon wave function,
where the $\gamma$ fluctuates into a virtual meson, predominantly a
$\rho^0$. The same framework as for $pp / p\overline{p}$ collisions can
then be recycled, with modest modifications.

To finish this section, a few words on theory and on MPI modelling in some
other Tevatron-era (and beyond) generators.

Generally, MPI ideas were gradually becoming more accepted.
An interesting (partial) alternative was raised by the CCFM equations
\cite{Ciafaloni:1987ur,Catani:1989sg}, which interpolate between the
DGLAP \cite{Gribov:1972ri,Altarelli:1977zs,Dokshitzer:1977sg} and
BFKL \cite{Kuraev:1977fs,Balitsky:1978ic} ones. Already BFKL allows
$p_{\perp}$-unordered evolution chains, and with CCFM such a behaviour
can be extended to higher $p_{\perp}$ scales. As illustrated in the LDC
model \cite{Andersson:1995ju}, this can then give what looks like several
(semi)hard interactions within one single chain.

\textsc{IsaJet} remained in use, even if slowly losing ground, with an
essentially unchanged description of the underlying event.

When \textsc{Herwig} was extended to hadronic collisions
\cite{Marchesini:1987cf}, it used an UE model/parametrization based on
the UA5 MB generator \cite{Alner:1986is}, which is purely soft physics.
MPIs were never made part of the Fortran \textsc{Herwig} core code.
Instead the UA5-based default could be replaced by the separate
\textsc{Jimmy} \cite{Butterworth:1996zw} add-on. Its basic ideas resemble
the ones in \textsc{Pythia}, but with several significant differences.
The impact-parameter profile is given by the electromagnetic form factor,
and at each given $b$ the number of MPIs (in addition to the hard process
itself) is given by a Poissonian
with an average proportional to the convolution of two form factors.
These MPIs are unordered in $p_{\perp}$, and all use unmodified PDFs.
Instead interactions that break energy-momentum conservation are rejected.
To handle beam remnants, it is assumed that each ISR shower initiator,
except the first, is a gluon; if not an additional ISR branching is made
to ensure this. Thereby it is possible to chain each MPI to the next in
colour, such that the remnant flavour structure is related only to
the first interaction. This handling allows all MPIs to be associated
with ISR and FSR, unlike \textsc{Pythia} at the time. Note that
\textsc{Jimmy} was intended for UE studies, and that
\textsc{Herwig+Jimmy} did not offer
an MB option. Such a framework was developed \cite{Borozan:2002fk}
but the code for it was never made public. Only with the C++ version
\cite{Bellm:2015jjp} did MPIs become a fully integrated part of the
core code, for UE and MB \cite{Bahr:2008dy,Gieseke:2016fpz}.

Another (later) multipurpose generator entrant is \textsc{Sherpa}
\cite{Gleisberg:2008ta}, which so far has based itself on the
\textsc{Pythia} MPI framework, but a new separate MPI model is under
development \cite{Martin:2012nm}.

Many generators geared towards minimum-bias physics also adapted
semihard MPI ideas. Notably, generators based on eikonalization
procedures typically already had contributions for soft and diffractive
MPI-style physics, and could add a further contribution for hard MPIs.
This means that a nondiffractive event can contain variable numbers of
soft $p_{\perp} = 0$ and hard $p_{\perp} > p_{\perp\mathrm{min}}$ MPIs.
Typically a Gaussian $b$ dependence is used, not necessarily with the
same width for all contributions. Main examples of such programs are
\textsc{DTUjet} \cite{Aurenche:1991pk,Aurenche:1994ev},
\textsc{PhoJet} \cite{Engel:1994vs,Engel:1995yda},
\textsc{DPMjet} \cite{Bopp:2005cr},
\textsc{Sibyll} \cite{Fletcher:1994bd,Ahn:2009wx},
\textsc{EPOS}   \cite{Werner:2005jf,Pierog:2013ria},
see also \cite{CONTRIB:MC-EPOS-HM},
and \textsc{QGSjet} \cite{Ostapchenko:2005nj,Ostapchenko:2010vb}.
It would carry too far to go into the details of these programs. Some
of them are in use at the LHC, and describe minimum-bias data quite
successfully. They are not only used for $pp$ collisions but often also
for $pA$ and $AA$, and for cosmic-ray cascades in the atmosphere.

\section{Multiparton PDFs and beam remnants}
\label{tssec:multiPDFs}

In 2004 the \textsc{Pythia} MPI model was significantly upgraded
\cite{Sjostrand:2004pf}, specifically to allow a more realistic
description of multiparton PDFs and beam remnants. Then ISR and FSR
could also be included for each MPI, not only the hardest one.

To extend the PDF framework, it is assumed that quark distributions
can be split into one valence and one sea part. In cases where this is
not explicit in the PDF parametrizations, it is assumed that the sea
is flavour-antiflavour symmetric, so that one can write \textit{e.g.}
\begin{equation}
u(x,Q^2) = u_{\mathrm{val}}(x,Q^2) + u_{\mathrm{sea}}(x,Q^2) =
u_{\mathrm{val}}(x,Q^2) + \overline{u}(x,Q^2).
\end{equation}
The parametrized $u(x, Q^2)$ and $\overline{u}(x, Q^2)$ distributions
can then be used to find the relative probability for a kicked-out
$u$ quark to be either valence or sea.

For valence quarks two effects should be considered. One is the
reduction in content by previous MPIs: if a $u$ valence quark has
been kicked out of a proton then only one remains, and if two then
none remain. In addition the constraint from momentum conservation
should be included, as already introduced in
Eq.~(\ref{tseq:simplefrescale}). Together this gives
\begin{equation}
u_{i,\mathrm{val}}(x,Q^2)
= \frac{N_{u,\mathrm{val,remain}}}{N_{u,\mathrm{val,original}}} \,
\frac{1}{X} \, u_{\mathrm{val}} \left(\frac{x}{X},Q^2\right)
~~\mathrm{with}~~ X = 1 - \sum_{j = 1}^{i - 1}x_j,
\label{tseq:valfrescale}
\end{equation}
for the $u$ quark in the $i$'th MPI, and similarly for the $d$. The
$1/X$ prefactor ensures that the $u_i$ integrates to the remaining
number of valence quarks. The momentum sum is also preserved, except
for the downwards rescaling for each kicked-out valence quark. The
latter is compensated by an appropriate scaling up of the gluon and
sea PDFs.

When a sea quark (or antiquark) $q_{\mathrm{sea}}$ is kicked out of a
hadron, it must leave behind a corresponding antisea parton in the
beam remnant, by flavour conservation, which can then participate in
another interaction. We can call this a companion
antiquark, $\overline{q}_{\mathrm{cmp}}$. In the perturbative
approximation the pair comes from a gluon branching
$g \to q_{\mathrm{sea}} + \overline{q}_{\mathrm{cmp}}$.
This branching often would not be in the perturbative
regime, but we choose to make a perturbative ansatz, and also to
neglect subsequent perturbative evolution of the $q_{\mathrm{cmp}}$
distribution. Even if approximate, this procedure should catch
the key feature that a sea quark and its companion should not be
expected too far apart in $x$ (or, better, in $\ln x$). Given a
selected $x_{\mathrm{sea}}$, the distribution in
$x = x_{\mathrm{cmp}} = y - x_{\mathrm{sea}}$ then is
\begin{eqnarray}
q_{\mathrm{cmp}}(x;x_{\mathrm{sea}}) & = & C \int_0^1
g(y) \, P_{g \to q_{\mathrm{sea}}\overline{q}_{\mathrm{cmp}}}(z)
\, \delta(x_{\mathrm{sea}} - zy) \, \mathrm{d}z
\nonumber \\
& = & C~\frac{g(x_{\mathrm{sea}}+x)}{x_{\mathrm{sea}}+x} \,
P_{g \to q_{\mathrm{sea}}\overline{q}_{\mathrm{cmp}}}
\left(\frac{x_{\mathrm{sea}}}{x_{\mathrm{sea}}+x}\right) ~.
\end{eqnarray}
Here $P_{g \to q\overline{q}}(z)$ is the standard DGLAP branching kernel,
$g(y)$ an approximate gluon PDF, and $C$ gives an overall normalization
of the companion distribution to unity.
Furthermore an $X$ rescaling is necessary as for valence quarks.
The addition of a companion quark does break the momentum sum rule,
this times upwards, and so is compensated by a scaling down of the
gluon and sea PDFs.

In summary, in the downwards evolution, the kinematic limit is
respected by a rescaling of $x$, as before. In addition the number
of remaining valence quarks and new companion quarks is properly
normalized. Finally, the momentum sum is preserved by a scaling of
gluon and (non-companion) sea quarks. All of these scalings should
not be interpreted as a physical change of the beam hadron, but merely
as reflecting an increasing knowledge of its contents, akin to
conditional probabilities.

At the end of the MPI + ISR generation sequence, a set of initiator
partons have been taken out of the beam, \textit{i.e.}\ partons that initiate
the ISR chains that stretch in to the hard interactions. The beam remnant
contains a number of leftover valence and companion quarks that carry
the relevant flavour quantum numbers, plus gluons and sea to make up the
total momentum. The latter are not book-kept explicitly, except for the
rare case when the remnant contains no valence or companion quarks at all,
and where a gluon is needed to carry the leftover momentum.

When the initiators are taken out of the incoming beam particle
they are assumed to have a primordial $k_{\perp}$. Naively this would be
expected to be of the order of the Fermi motion inside the proton,
\textit{i.e.}\ a few hundred MeV. In order to describe the low-$p_{\perp}$
tail of the $Z^0$ spectrum a rather higher value of the order of 2~GeV
seems to be required. This suggests imperfections in the modelling of ISR
at small scales, specifically how and at what $p_{\perp}$ scales it should
be stopped. Also note that ISR dilutes the $k_{\perp}$ at the hard
interaction by a factor $x_{\mathrm{hard}} / x_{\mathrm{init}}$,
\textit{i.e.}\ by the fraction of the initiator longitudinal momentum
that reaches the hard interaction. More ISR means a higher
$x_{\mathrm{init}}$ and thus more dilution, counteracting the $p_{\perp}$
gained by the ISR itself.

Given such considerations, a Gaussian distribution is used for the
primordial $k_{\perp}$, with a width that depends on the scale $Q$
($p_{\perp}$) of each MPI, increasing smoothly from 0.36~GeV (= the
string hadronization $p_{\perp}$) at small $Q$ to 2~GeV at large.
There is also a check that the $k_{\perp}$ does not become too big
for a low-mass system. The beam remnants are each given a $k_{\perp}$
at the lower scale, but in addition they collectively have to take the
recoil to ensure that the net $p_{\perp}$ vanishes among the initiators
and remnants.

The beam remnants also share leftover energy and longitudinal momentum.
This is done by an ansatz of specific $x$ spectra for valence quarks,
valence diquarks, and companion quarks. The $x$ values determine the
relative fractions partons take of the lightcone momentum $E \pm p_z$,
with $+$ ($-$) for the beam moving in the $+z$ ($-z$) direction.
It is not possible to fully conserve four-momentum inside each
remnant + initiators system individually, however. Actually, by their
relative motion the beam remnants together obtain a spectrum of
invariant masses stretching well above the proton mass. Instead overall
energy and momentum is preserved by longitudinal boosts of the two
remnant subsystems, which effectively corresponds to a shuffling of
four-momentum between the two sides. It is possible to generate too big
remnant masses, but usually this can be fixed by a reselection of
remnant $x$ values.

\begin{figure}[t]
\centerline{\includegraphics[width=80mm]{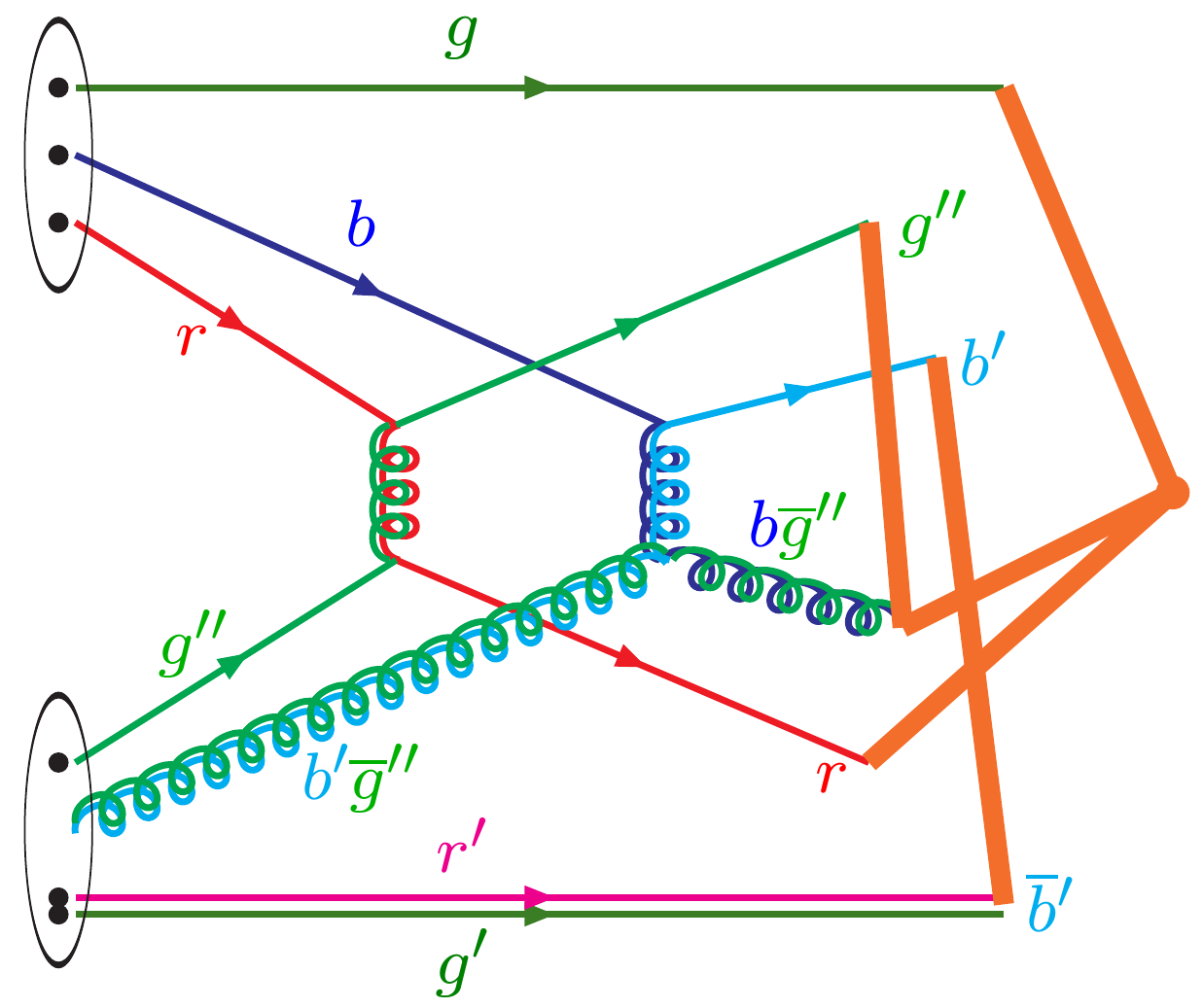}}
\caption{Example of an event where two valence quarks are kicked out
from a proton, giving a junction topology. A possible colour flow
is indicated, where primed colours are distinguishable from unprimed ones
in the $N_C \to \infty$ limit. The remnant diquark is bookkept as a unit
with $r' + g' = \overline{b}'$. Thick orange lines indicate strings
stretched between outgoing partons, with the junction placed rightmost
to avoid clutter.}
\label{tsfig:junctioncolours}
\end{figure}

What remains to consider is how the colours of partons are connected with
each other to give the string pieces that eventually hadronize. This was
one of the key stumbling blocks in the original model, especially
what to do if several valence quarks are kicked out of a proton,
Fig.~\ref{tsfig:junctioncolours}.
The main new tool at our disposal at this point is an implementation of
junction fragmentation \cite{Sjostrand:2002ip}. A junction is a vertex at
which three string pieces come together, in a Y-shaped topology, and with
each string stretching out to a quark, in the simplest case. The net baryon
number then gets to be associated with the junction: given enough energy
each string piece can break by the production of new $q\overline{q}$
(or $qq\overline{q}\overline{q}$) pairs, splitting off mesons (or
baryon--antibaryon pairs), leaving the innermost $q$ on each string
to form a baryon together. An antijunction carries a negative baryon
number, since the three strings in this case stretch out to antiquarks.

The rest frame of the junction is obtained in a symmetric configuration,
where the opening angle between any pair of outgoing string ends is
$2\pi/3 = 120^{\circ}$. This also defines the approximate rest frame
of the central baryon. In cases where only one quark is kicked out of
an incoming proton, the remaining two quarks in the beam remnant have
a tiny opening angle in the collision rest frame, meaning the junction
is strongly boosted in the forward direction, along with the two quarks,
and these can then together be treated as a single unresolved diquark.
If two valence quarks are kicked out, however, the junction can end up
far away from the beam remnant itself. Note that a junction is normally
not associated with the original quarks after a collision, owing to
colour exchange.

A major complication is that the three strings may be stretched via
various intermediate gluons out to the (new) endpoint quarks, and then
the string motion and fragmentation becomes far more complex. It is such
general issues that had taken time to resolve, at least approximately.
Also systems containing a junction and an antijunction connected to
each other need to be described.

Colours can be traced within each MPI individually, both through the
hard interaction and the related ISR and FSR cascades, in the
$N_C \to \infty$ limit \cite{tHooft:1973alw}. If this limit is taken
seriously, however, the beam remnants have to compensate the colours of
all initiator partons, which means that they build up a
high colour charge, which has to be carried by an unrealistically large
number of remnant partons. It is more plausible, although a bit extreme
in the other direction, to assume that the colour taken by
one initiator is the anticolour of another one. It is such a strategy that
allows us to work with the minimal number of beam remnants that preserves
net flavour (or a single gluon if all valence quarks are kicked out).
A sea quark initiator can be associated with its companion antiquark,
be that another initiator or a remnant parton, and together be traced
back to an imagined gluon that can be attached as above.

Thus only the valence colours remain. A proton can be described as a
quark plus a diquark if none or one valence quark is kicked out, else as
three quarks in a junction topology. It is along these original colour
lines that the gluon initiators are attached one by one. Three main
alternatives are implemented for the order of these attachments, from
completely random to ones that favour smaller string lengths $\lambda$.
(These connections can give a gluon the same colour as anticolour,
which clearly is unphysical. Such colour associations are rejected and
others tried.) Not even in the latter case does
$\langle p_{\perp} \rangle (n_{\mathrm{ch}})$ rise as fast as observed in
the data, however, which suggests that a mere arrangement of colours
in the initial state is not enough. A mechanism is also needed for CR
in the final state, as already used in the earlier models.

Again the $\lambda$ measure is used to pick such reconnections: a string
piece $ij$ stretched between partons $i$ and $j$ and another $mn$
between $m$ and $n$ can reconnect to $in$ and $mj$ if
$\lambda_{in} + \lambda_{mj} < \lambda_{ij} + \lambda_{mn}$. A free
strength parameter is introduced to regulate the fraction of pairs
that are being tested this way. With this further mechanism at hand
it now again becomes possible to describe
$\langle p_{\perp} \rangle (n_{\mathrm{ch}})$ data approximately.

\section{Interleaved evolution}
\label{tssec:interleave}

In models up until now, MPIs have been considered one by one. Once an
MPI has been picked, the ISR and FSR associated with it has been
generated before moving on to the next. This ordering is not trivial,
since both the MPI and ISR mechanisms need to take momentum from the beam
remnants, and therefore are in direct competition. If instead all MPIs
had been generated first, and all ISR added only afterwards, the number
of MPIs would have been higher and the amount of ISR less.

Time ordering does not give any clear guidance what is the correct
procedure. We have in mind a picture where all MPIs happen simultaneously
at the collision moment, while ISR stretches backwards in time from it,
and FSR forwards. But we have no clean way of separating the hard
interactions themselves from the virtual ISR cascades that ``already''
exist in the colliding hadrons.

Instead we choose the same guiding principle as we did when we originally
decided to consider MPIs ordered in $p_{\perp}$: it is most important
to get the hardest part of the story ``right'', and then one has to live
with an increasing level of approximation for the softer steps.
With the introduction of $p_{\perp}$-ordered showers \cite{Sjostrand:2004ef}
it became possible to choose $p_{\perp}$ as common evolution scale.
Initially only MPI and ISR were interleaved, with FSR left to the end.
This caught the important momentum competition between MPI and ISR,
so was the big step. When \textsc{Pythia}~8 was written
\cite{Sjostrand:2007gs} full MPI/IRS/FSR interleaving \cite{Corke:2010yf}
was default from the beginning. Going straight for the latter formulation,
the scheme is characterized by one master formula
\begin{eqnarray}
\frac{\mathrm{d} \mathcal{P}}{\mathrm{d} p_{\perp}}&=&
\left( \frac{ \mathrm{d}\mathcal{P}_{\mathrm{MPI}}}{\mathrm{d} p_{\perp}}  +
\sum   \frac{ \mathrm{d}\mathcal{P}_{\mathrm{ISR}}}{\mathrm{d} p_{\perp}}  +
\sum   \frac{ \mathrm{d}\mathcal{P}_{\mathrm{FSR}}}{\mathrm{d} p_{\perp}} \right)
\nonumber \\
 & \times & \exp \left( - \int_{p_{\perp}}^{p_{\perp\mathrm{max}}}
\left( \frac{ \mathrm{d}\mathcal{P}_{\mathrm{MPI}}}{\mathrm{d} p_{\perp}'}  +
\sum   \frac{ \mathrm{d}\mathcal{P}_{\mathrm{ISR}}}{\mathrm{d} p_{\perp}'}  +
\sum   \frac{ \mathrm{d}\mathcal{P}_{\mathrm{FSR}}}{\mathrm{d} p_{\perp}'}
\right) \mathrm{d} p_{\perp}' \right)
\label{tseq:combinedevol}
\end{eqnarray}
that probabilistically determines what the next step will be.
Here the ISR sum runs over all incoming partons, two per
already produced MPI, the FSR sum runs over all outgoing partons,
and $p_{\perp\mathrm{max}}$ is the $p_{\perp}$ of the previous step.
Starting from the hardest interaction, Eq.~(\ref{tseq:combinedevol})
can be used repeatedly to construct a complete parton-level event.

Since each of the three terms contains a lot of complexity, with
matrix elements, splitting kernels and PDFs in various combinations,
it would seem quite challenging to pick a $p_{\perp}$ according to
Eq.~(\ref{tseq:combinedevol}). Fortunately the ``winner-takes-it-all''
trick (which is exact \cite{Corke:2009tk}) comes to the rescue. In it
you select a $p_{\perp\mathrm{MPI}}$ value as if the other terms did not
exist in the equation, and correspondingly a $p_{\perp\mathrm{ISR}}$ and a
$p_{\perp\mathrm{FSR}}$. Then the one of the three that is largest decides
what is to come next. Inside the ISR and FSR sums one can repeat the
same trick, \textit{i.e.}\ only consider one term at a time and decide
which term gives the highest $p_{\perp}$.

The multiparton PDFs introduced in Sec.~\ref{tssec:multiPDFs}
play a key role, to help select a new MPI or perform ISR on an already
existing one. Note that momentum and flavour should not be deducted for
the current MPI itself when doing ISR. To exemplify, if the valence $d$
quark has been kicked out of a proton in a given MPI, then there are
no such $d$'s left for other MPIs, neither in ISR nor in MPI steps,
but for the given MPI a valence $d$ at higher $x$ is still available
as a potential mother to the current $d$.

In summary, $p_{\perp}$ fills the function of a kind of factorization
scale, where the perturbative structure above it has been resolved,
while the one below it is only given an effective description
\textit{e.g.}\ in terms of multiparton PDFs. A decreasing $p_{\perp}$
scale should then be viewed as an evolution towards increasing resolution;
given that the event has a particular structure when viewed at some
$p_{\perp}$ scale, how might that picture change when the resolution
cutoff is reduced by some infinitesimal $\mathrm{d} p_{\perp}$? That is,
let the ``harder'' features of the event set the pattern to which
``softer'' features have to adapt.

\begin{figure}[t]
\centerline{\includegraphics[width=110mm]{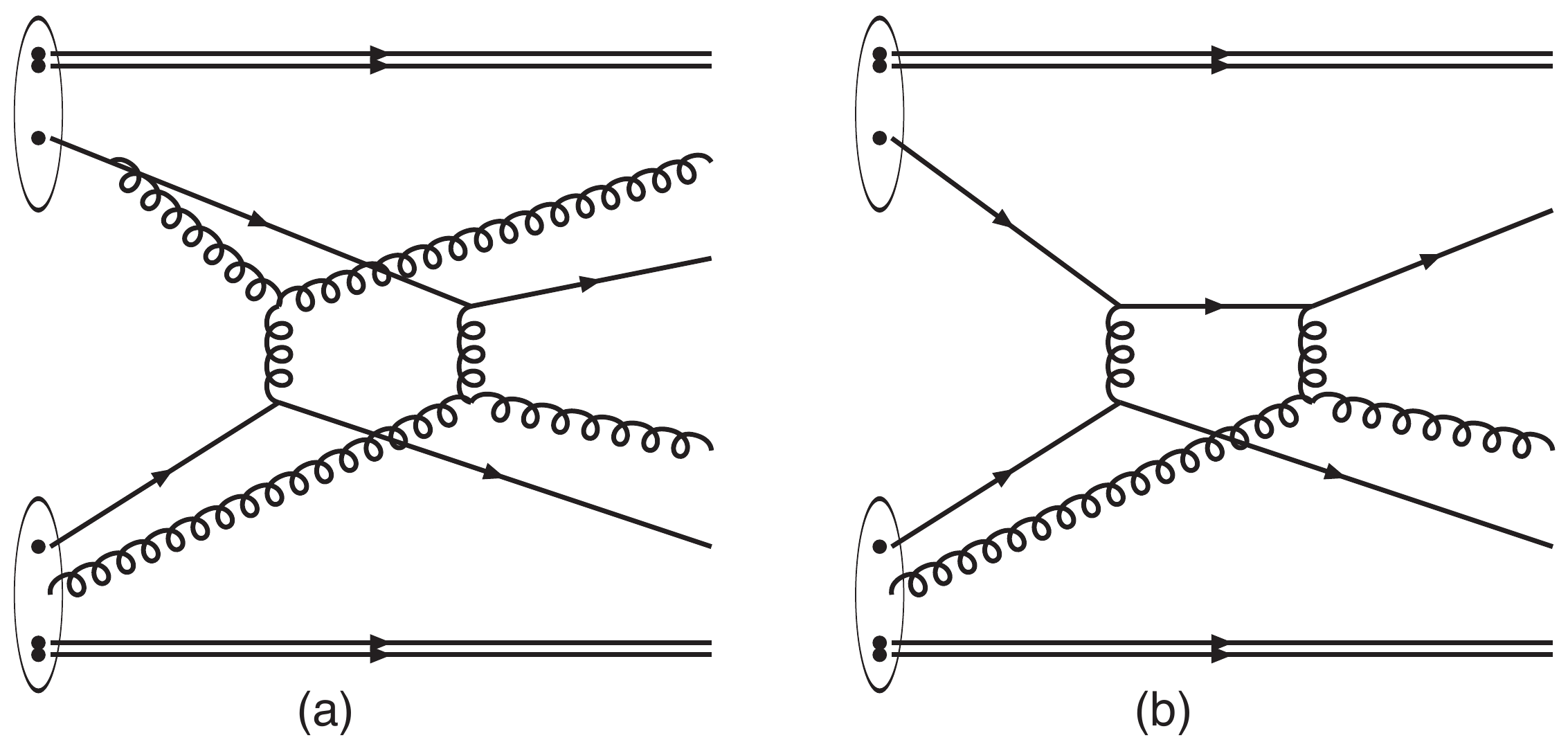}}
\caption{(a) Joined interactions. (b) Rescattering.}
\label{tsfig:intertwine}
\end{figure}

\section{Intertwined evolution}
\label{tssec:intertwine}

The above interleaving introduces a strong but indirect connection between
different MPIs, in that each parton still has a unique association with
exactly one MPI and its associated ISR and FSR. But this is likely not
the full story; there are several ways in which the different MPIs may
be much closer intertwined
\cite{Blok:2011bu,Diehl:2011yj,Gaunt:2012dd,Diehl:2017kgu},
see also \cite{CONTRIB:TH-DPS-INTRO}. The complexity then is
significantly increased, and none of these further mechanisms are
included by default in \textsc{Pythia}, but some have been studied
and partly implemented.

The first possibility is joined interactions (JI) \cite{Sjostrand:2004ef},
Fig.~\ref{tsfig:intertwine}(a).
In it two partons participating in two separate MPIs may turn out to have
a common ancestor when the backwards ISR evolution traces their prehistory.
The joined interactions are well-known in the context of the forwards
evolution of multiparton densities \cite{Konishi:1979cb,Kirschner:1979im,%
Shelest:1982dg,Snigirev:2003cq,Korotkikh:2004bz}. It can approximately be
turned into a backwards evolution probability for a branching $a \to bc$
\begin{equation}
\mathrm{d} \mathcal{P}_{bc}(x_b, x_c, Q^2) \simeq
\frac{\mathrm{d} Q^2}{Q^2} \, \frac{\alpha_{\mathrm{s}}}{2 \pi} \
\frac{x_a f_a(x_a, Q^2)}{x_b f_b(x_b, Q^2) \, x_c f_c(x_c, Q^2)} \,
z (1-z) P_{a \to bc}(z) ~,
\label{tseq:fcorrevol}
\end{equation}
with $x_a = x_b + x_c$ and $z = x_b/(x_b + x_c)$. The main approximation
is that the two-parton differential distribution has been been factorized
as $f^{(2)}_{bc}(x_b, x_c, Q^2) \simeq f_b(x_b, Q^2) \, f_c(x_c, Q^2)$
to put the equation in terms of more familiar quantities.

Just like for the other processes considered, a form factor is given by
integration over the relevant $Q^2$ range and exponentiation. Associating
$Q \simeq p_{\perp}$, as is already done for normal ISR,
Eq.~(\ref{tseq:fcorrevol}) can be turned into a
$\sum \mathrm{d} \mathcal{P}_{\mathrm{JI}} / \mathrm{d}p_{\perp}$ term
that can go into Eq.~(\ref{tseq:combinedevol}) along with the other three.
The JI sum runs over all pairs of initiator partons with allowable flavour
combinations, separately for the two incoming hadrons. A gluon line can
always be joined with a quark or another gluon one, and a sea quark and
its companion can be joined into a gluon. The parton densities are defined
in the same spirit as before, \textit{e.g.}\ $f_b(x_b,p_{\perp}^2)$ and
$f_c(x_c,p_{\perp}^2)$ are squeezed to be smaller than $X$, where $X$ is
reduced from unity by the momentum carried away by all but the own
interaction, and for $f_a(x_a,p_{\perp}^2)$ by all but the $b$ and $c$
interactions.

Technical complications arise when the kinematics of JI branchings has
to be reconstructed, notably in transverse momentum,  and the code to
overcome these was never written. The reason is that already the evolution
itself showed that JI effect are small. Most events do not contain
any JI at all above the ISR cutoff scale, and in those that do the JI
tends to occur at a low $p_{\perp}$ value. There are two reasons for this.
One is numerics: the number of parton pairs that can be joined increases
as more MPIs have already been generated. The other is the PDF behaviour:
for all but the smallest $Q^2 = p_{\perp}^2$ scales the huge number of
small-$x$ gluons and sea quarks dominate, and it is only close to the
lower evolution cutoff that the few valence quarks and high-$x$ gluons
play an increasingly important role in the ISR backwards evolution.

The second intertwining possibility is rescattering, \textit{i.e.}\ that
a parton from one incoming hadron consecutively scatters against two or
more partons from the other hadron, Fig.~\ref{tsfig:intertwine}(b).
The simplest case, $3 \to 3$, \textit{i.e.}\ one rescattering, has been
studied \cite{Paver:1982yp,Paver:1983hi,Paver:1984ux}.
The conclusion is that it should be less important than two separated
$2 \to 2$ processes: $3 \to 3$ and $2 \times (2 \to 2)$ contain the same
number of vertices and propagators, but the latter wins by involving one
parton density more. The exception could be large $p_{\perp}$ and $x$ values,
but there $2 \to n, n \geq 3$  QCD radiation anyway is expected to be the
dominant source of multijet events.

For rescattering, a detailed implementation is available as an option
in \textsc{Pythia} \cite{Corke:2009tk}, as follows.
Previously we have described how partons are taken out of the conventional
PDFs after each MPI (and ISR), such that less and less of the original
momentum remains. If we now should allow a rescattering then a scattered
parton has to be put back into the PDF, but now as a $\delta$ function.
It can be viewed as a quantum mechanical measurement of the wave function
of the incoming hadron, where the original ``squared wave function''
$f(x,Q^2)$ in part collapses by the measurement process of one of the
partons in the hadron. That is, one degree of freedom has now
been fixed, while the remaining ones are still undetermined.
A hadron can therefore be characterized by a new PDF
\begin{equation}
f(x,Q^2) \rightarrow
f_{\mathrm{rescaled}}(x,Q^2) + \sum_{i} \delta(x - x_i)
= f_{\mathrm{u}}(x,Q^2) + f_{\delta}(x,Q^2) ~,
\end{equation}
where $f_{\mathrm{u}}$ represents the unscattered part of the hadron and
$f_{\delta}$ the scattered one. The scattered partons have the same
$x$ values as originally picked, in the approximation that small-angle
$t$-channel gluon exchange dominates, but more generally there will be
shifts. The sum over delta functions runs over all partons that are
available to rescatter, including outgoing states from hard or MPI
processes and partons from ISR or FSR branchings. All the partons of
this disturbed hadron can scatter, and so there is the possibility for
an already extracted parton to scatter again.

With the PDF written in this way, the original MPI probability
in Eq.~(\ref{tseq:combinedevol}) can now be generalised to
include the effects of rescattering
\begin{equation}
\frac{\mathrm{d} \mathcal{P}_{\mathrm{MPI}}}{\mathrm{d} p_{\perp}} \rightarrow
\frac{\mathrm{d} \mathcal{P}_{\mathrm{uu}}}{\mathrm{d} p_{\perp}} +
\frac{\mathrm{d} \mathcal{P}_{\mathrm{u \delta}}}{\mathrm{d} p_{\perp}} +
\frac{\mathrm{d} \mathcal{P}_{\mathrm{\delta u}}}{\mathrm{d} p_{\perp}} +
\frac{\mathrm{d} \mathcal{P}_{\mathrm{\delta \delta}}}{\mathrm{d} p_{\perp}}
\label{tseq:combinedresc}
~,
\end{equation}
where the uu component represents the original MPI probability, the
$\mathrm{u \delta}$ and $\mathrm{\delta u}$ components a single
rescattering and the $\mathrm{\delta \delta}$ component a double
rescattering. In this way, rescattering interactions are included
in the common $p_{\perp}$ evolution of MPI, ISR and FSR.

Again the evolution in $p_{\perp}$ should be viewed as a resolution ordering.
In a time-ordered sense, a parton could scatter at a high $p_{\perp}$ scale
and rescatter at a lower one, or the other way around, with comparable
probabilities. To simplify the already quite complicated machinery, it is
chosen to set up the generation as if the rescattering occurs both at a
lower $p_{\perp}$ and a later time than the ``original'' scattering,
while retaining the full rescattering rate.

In simple low-angle scatterings there is a unique assignment of
scattered partons to either of the two colliding hadrons
$A$ and $B$, but for the general case this is not so simple. Different
prescriptions have been tried, including the most extreme where all
scattered partons can scatter again against partons from either beam.
It turns out that differences are small for single rescatter. The
scheme dependence is bigger for double rescattering, but this process
is anyway significantly suppressed relative to the single rescatters.

The real problem, however, is how to handle kinematics when ISR, FSR and
primordial $k_{\perp}$ are added to rescattering systems. To propagate
recoils between systems that are partly intertwined but also partly
separate requires what is the most nontrivial code in the whole
\textsc{Pythia} MPI framework, and it would carry too far to describe
it here.

More important are the results. At the LHC we predict that of the order
of half of all minimum-bias events contains at least one rescattering,
and for events with hard processes the fraction is even larger. Double
rescattering is rare, however, and can be neglected. Evaluating the
consequences of rescattering is still challenging, since it is a
secondary effect within the MPI machinery. The precise amount of MPI
has to be tuned to data, \textit{e.g.}\ by varying the $p_{\perp 0}$ turn-off
parameter, so the introduction of rescattering to a large extent can be
compensated by a slight decrease in the amount of ``normal'' $2 \to 2$
MPIs. Worse still, most MPIs are soft to begin with, and rescattering
then introduces a second scale, even softer than the ``original''
scattering. Thus, rescattering is typically associated with particle
production at the lower limit of what can be reliably detected. There
are some effects of the Cronin type \cite{Cronin:1974zm} in hadronic
$p_{\perp}$ spectra, \textit{i.e.}\ a shift towards higher $p_{\perp}$,
but too small to offer a convincing signal.

If one zooms in on the tail of events at larger scales, the rate of
\mbox{(semi-)hard} three-jet events from rescattering is of the same order
in $\alpha_{\mathrm{s}}$ as four-jets from DPS, but the background from
ISR and FSR starts one order earlier for three-jets. Furthermore, DPS
has some obvious characteristics to distinguish it from $2 \to 4$
radiation topologies: pairwise balanced jets with an isotropic relative
azimuthal separation. No corresponding unique kinematic features are
expected for $3 \to 3$ rescattering vs.\ $2 \to 3$ ISR/FSR, and we have
not found any either. Hopefully smarter people will one day find the
right observable.

The third and most dramatic intertwining possibility is that the
perturbative cascades grip into each other. An example is the ``swing''
mechanism, whereby two dipoles in the initial state can reconnect
colours, which is a key aspect of the \textsc{Dipsy} generator
\cite{Avsar:2006jy,Bierlich:2014xba}, see also \cite{CONTRIB:MC-DIPSY-HM}.
Currently there is no  mechanism of this kind in \textsc{Pythia}.

\section{An $x$-dependent impact-parameter profile}
\label{tssec:xdepb}

As described in Sec.~\ref{tssec:dampenimpact}, a double Gaussian
impact-parameter profile was the initial choice, and remained
the default for many years. With the introduction of full ISR/FSR
in all MPIs, Sec.~\ref{tssec:multiPDFs}, a single Gaussian is
almost giving enough fluctuations. The ``hot spots'', that the double
Gaussian was introduced to represent, now are obtained in part
by the ISR/FSR cascades associated with each MPI, with somewhat fewer
separate MPIs needed for the same overall activity. Put another way, the
varying ISR/FSR activity introduces another mechanism for fluctuations,
while a smaller $\langle n_{\mathrm{MPI}} \rangle$ means that the scaled
width $\sigma_{\mathrm{MPI}}(b) / \langle n_{\mathrm{MPI}}(b) \rangle$
goes up, both reducing the need for broader-than-Gaussian profiles.

Given the reduced sensitivity to non-Gaussian profiles, a one-parameter
alternative was therefore introduced
\begin{equation}
\widetilde{\mathcal{O}}(b) = \exp\left( -b^d \right) ~,
\end{equation}
where $d < 2$ gives more fluctuations than a Gaussian and $d > 2$ less.
Do note that the expression is for the overlap, not for the individual
hadrons, for which no simple analytical form is available.

One aspect, however, is that we have assumed the transverse $b$ profile
to be decoupled from the longitudinal $x$ one. This is not the expected
behaviour, because low-$x$ partons should diffuse out in $b$ during
the evolution down from higher-$x$ ones \cite{Frankfurt:2005mc},
see also \cite{CONTRIB:TH-GPDF}. Additionally, if $b = 0$ is defined
as the center of energy of a hadron, then by definition a parton with
$x \to 1$ also implies $b \to 0$.

Such diffusion was already studied in Sec.~\ref{tssec:interlude}.
In the later study \cite{Corke:2011yy}, described here, there was no
attempt to trace the evolution of cascades in $x$. Rather it is assumed
that the $b$ distribution of partons at any $x$ can be described by a
simple Gaussian, but with an $x$-dependent width:
\begin{equation}
  \rho(r, x) \propto \frac{1}{a^3(x)} \, \exp
                     \left( - \frac{r^2}{a^2(x)} \right)
~~\mathrm{with}~~
  a(x) = a_0 \left( 1 + a_1 \ln \frac{1}{x} \right) ~,
\end{equation}
with $a_0$ and $a_1$ to be determined. The overlap is then given by
\begin{equation}
\widetilde{\mathcal{O}}(b, x_1, x_2) = \frac{1}{\pi} \,
  \frac{1}{a^2(x_1) + a^2(x_2)} \,
  \exp \left( - \frac{b^2}{a^2(x_1) + a^2(x_2)} \right) ~.
\end{equation}
In principle one could argue that also a third length scale should be
included, related to the transverse distance the exchanged propagator
particle, normally a gluon, could travel. This distance should be made
dependent on the $p_{\perp}$ scale of the interaction. For simplicity
this further complication is not considered but, on the other hand,
a finite effective radius is allowed also for $x \to 1$. The $x \to 1$
limit is not much probed in MB, since the bulk of MPIs occur at small $x$,
but can be relevant for UE studies, \textit{e.g.}\ for the production of
new particles at high mass scales.

The two parameters $a_0$ and $a_1$ could be tuned at each CM energy.
One combination of them is fixed so as to reproduce $\sigma_{\mathrm{nd}}$.
If the wider profile of low-$x$ partons is related to the growth of
$\sigma_{\mathrm{nd}}(s)$, then $a_1$ can be constrained by the
requirement that $a_0$ should be independent of energy. This is
reasonably well fulfilled for $a_1 = 0.15$, which is therefore taken
as default in this scenario.

The generation of events is more complicated with an $x$-dependent
overlap, but largely involves the same basic principles.
Again the $b$ value of an event is selected in conjunction with
the kinematics of the hardest interaction, and is thereafter fixed.
The acceptance of subsequent MPIs is proportional to
$\widetilde{\mathcal{O}}(b, x_1, x_2)$, where $x_1$ and $x_2$ are the
values for the MPI under consideration.

In overall MB and UE phenomenology, the scenario with $a_1 = 0.15$ ends
up roughly halfway between the single and double Gaussian ones.
It depends on the process under consideration, however. A 10~GeV
$\gamma^*$ would give an MPI activity close to the single Gaussian,
while a 1~TeV $Z'$ would have markedly higher MPI activity, since
larger $x$ values would be accessed in the latter process.

Unfortunately, in the experimental tunes to MB and UE data that were made
to this model variant \cite{ATLAS:2012uec}, no convincing evidence were
found for an $a_1 > 0$. A dedicated study of how the UE varies as a
function of the mass of $\mu^+\mu^-$ pairs could provide the definitive
test.

\section{Diffraction}
\label{tssec:diffr}

The nature of diffraction is not obvious, and has been addressed from
different points of view over the years, see
\cite{CONTRIB:LOWX+DIFFRACTION}.
From the perspective of this article the story begins with the
Ingelman--Schlein model \cite{Ingelman:1984ns},
wherein it is assumed that the exchanged Pomeron
$\mathbb{P}$ can be viewed as a hadronic state, and that therefore the
creation of a diffractive system can be described as a hadron-hadron
collision. This implies that the $\mathbb{P}$ has PDFs, which enables
hard processes to occur, as was confirmed by the observation of jet
production in diffractive systems \cite{Bonino:1988ae}. The
\textsc{PomPyt} program \cite{Bruni:1993ym} combined the $\mathbb{P}$
flux inside the proton with the $\mathbb{P}$ PDFs, both largely
determined from HERA data, and used \textsc{Pythia} to produce complete
hadronic final states.

Originally \textsc{Pythia} itself had a more primitive description,
based on a purely longitudinal structure of the diffractive system,
for single diffraction either stretched between a kicked-out valence
quark and a diquark remnant, or stretched in a hairpin configuration
with a kicked-out gluon connected to both a quark and a diquark in the
remnant. This was sufficient for low-mass diffraction, but clearly not
for high-mass one. Therefore a complete MPI machinery was implemented
\cite{Navin:2010kk,Corke:2010yf}. The diffractive mass is selected
as a first step. Thereafter $\mathbb{P}$ and ordinary $p$ PDFs are used
to generate an ordinary sequence of MPIs.

There are some catches, however. One is that $\mathbb{P}$ PDFs often are
not normalized to unit momentum, in part based on theoretical arguments
but mainly because what hard processes probe is the convolution of
$\mathbb{P}$ flux and PDFs, not each individually. For the handling of
momentum conservation issues in a generator, however, it is essential
to let the MPI machinery have access to properly rescaled PDFs, as in
Sec.~\ref{tssec:multiPDFs}, and for that an implicit normalization to
unity is used. Another is that the MPI machinery also requires a
normalization to an effective $\sigma_{\mathbb{P} p}$, to replace the
$\sigma_{\mathrm{nd}}$ used for normal nondiffractive events, starting
from Eq.~(\ref{tseq:nMPIavg}). Again this is not directly measurable,
so a default value of 10~mb has been chosen to give $\mathbb{P}p$
properties, such as $\langle n_{\mathrm{ch}} \rangle$, comparable with
$pp/p\overline{p}$ at the same mass.

The diffractive machinery generates a mass spectrum that stretches
down to $\sim 1.3$~GeV, but obviously it is not possible to apply an
MPI philosophy that low. Therefore, below 10~GeV, the original
longitudinal description is recovered. At low masses the kicked-out
valence quark scenario is assumed to dominate, but then fall off in
favour of a kicked-out gluon. Above 10~GeV there is a transition region,
wherein the MPI description gradually takes over from the longitudinal
one.

In its details several further deliberations and parameters are involved.
There is also a somewhat separate MBR model\cite{Ciesielski:2012mc} as
an option.

Diffraction raises more MPI questions. In the collision of two incoming
protons, one $\mathbb{P}$ exchange could imply a diffractive topology with
a rapidity gap, but other simultaneously occurring MPIs would fill in that
gap and produce an ordinary MB event. A spectacular example is Higgs
production by gauge-boson fusion, $W^+W^- \to H^0$ and $Z^0Z^0 \to H^0$,
where the naive process should result in a large central gap only
populated by the Higgs decay products, since no colour exchange is
involved. Including MPIs, this gap largely fills up
\cite{Dokshitzer:1991he}, although a fraction of the events should
contain no further MPIs \cite{Bjorken:1992er}, a fraction denoted as
the Rapidity Gap Survival Probability. Such a picture has been given
credence by the observation of ``factorization breaking'' between HERA
and the Tevatron: the $\mathbb{P}$ flux and PDFs determined at HERA predict
about an order of magnitude more QCD jet production than observed at the
Tevatron \cite{Affolder:2000vb}.

In a recent study \cite{Rasmussen:2015qgr} a dynamical description
of such factorization breaking is implemented, as a
function of the hard-process kinematics, to predict the
resulting event structure for hard diffraction in hadronic collisions.
This is done in three steps. Firstly, given a hard process which has
been selected based on inclusive PDFs, the fraction of a PDF that
should be associated with diffraction is calculated, obtained as a
convolution of the $\mathbb{P}$ flux and its PDFs. Secondly, the full
MPI framework of \textsc{Pythia}, including also ISR and FSR effects,
is applied to find the fraction of events without any further MPIs.
Those events that survive these two steps define the diffractive event
fraction, while the rest remain as regular nondiffractive events.
Thirdly, diffractive events may still have MPIs within the $\mathbb{P}p$
subsystem, and therefore the full hadron-hadron underlying-event
generation machinery is repeated for this subsystem. The nondiffractive
events are kept as they are in this step.

For typical processes, such as QCD jets or $Z^0$ production, the PDF
step gives $\sim$10\% of the events to be of diffractive nature.
The requirement of having no further MPIs gives an approximate factor
of 10 further suppression, so that only around 1\% of the processes
show up as diffraction. There numbers are in overall agreement with
the experimental ones, but the availability of a complete implementation
should allow more detailed tests. Unfortunately there is still a
non-negligible uncertainty, both for the model itself and for the
external input, such as $\mathbb{P}$ PDFs.

\section{Colour reconnection}
\label{tssec:colrec}

Colour reconnection is a research field on its own, although tightly
connected to MPIs, and has been reviewed elsewhere
\cite{Christiansen:2015yca,Sjostrand:2017ele}. Here only some brief
notes follow, again with emphasis on \textsc{Pythia} aspects.

As mentioned in Sec.~\ref{tssec:dampenimpact}, CR was essential to
obtain a rising $\langle p_{\perp} \rangle (n_{\mathrm{ch}})$, and that
has remained a constant argument over the years, still valid today:
separate MPIs must be colour-connected in such a way that topologies
with a reduced $\lambda$ measure, Eq.~(\ref{tseq:lambda}), are favoured.

LEP~2 offered a good opportunity to search for CR effects.
Specifically, in a process
$e^+ e^- \to W^+ W^- \to q_1 \overline{q}_2 q_3 \overline{q}_4$,
CR could lead to the formation of alternative ``flipped'' singlets
$q_1 \overline{q}_4$ and $q_3\overline{q}_2$, and correspondingly
for more complicated parton/string topologies. Such CR would be
suppressed at the perturbative level,
since it would force some $W$ propagators off the mass shell. This
suppression would not apply in the soft region. Two main models were
developed in the \textsc{Pythia} context \cite{Sjostrand:1993hi}.
Strings are viewed as elongated bags in scenario I, and reconnection is
proportional to the space--time overlap of these bags. In scenario II,
strings are instead imagined as vortex lines, and two cores need to cross
each other for a reconnection to occur. In either case it is additionally
possible to allow only reconnections that reduce $\lambda$, scenarios
I$'$ and II$'$. Based on a combination of results from all four LEP
collaborations, the no-CR null hypothesis is excluded at 99.5\% CL
\cite{Schael:2013ita}. Within scenario I the best description is obtained
for $\sim$50\% of the 189~GeV $W^+W^-$ events being reconnected, in
qualitative agreement with the \textsc{Pythia} predictions.

Unfortunately it is more difficult to formulate a similarly detailed
model of the space--time picture of hadronic collisions, and this has
not been done for \textsc{Pythia}. (In part such pictures are presented
\textit{e.g.}\ in \textsc{Dipsy} and \textsc{Epos}.) Instead simpler
scenarios for reducing the $\lambda$ have been used. In total
\textsc{Pythia}~6 came to contain twelve models, many of them involving
annealing strategies.

The current \textsc{Pythia}~8 \cite{Sjostrand:2014zea} initially only
contained one model, which still is the default. In it two MPIs can be
merged with a probability
$\mathcal{P} = r^2 p_{\perp 0}^2 / (r^2 p_{\perp 0}^2 + p_{\perp\mathrm{lower}}^2)$,
where $r$ is a free parameter, $p_{\perp 0}$
is the standard dampening scale of MPIs, and $p_{\perp\mathrm{lower}}$
is the scale of the lower-$p_{\perp}$ MPI.  Each gluon of the latter MPI
is put where it increases $\lambda$ the least for the higher-$p_{\perp}$ MPI.
The procedure is applied iteratively, so for any MPI the probability
of being reconnected is
$\mathcal{P}_{\mathrm{tot}} = 1 - (1 - \mathcal{P})^{n_{>}}$,
where $n_{>}$ is the number of MPIs with higher $p_{\perp}$.

A new QCD-based CR model \cite{Christiansen:2015yqa} implements a
further range of reconnection possibilities, notably allowing the
creation of junctions by the fusion of two or or three strings.
The relative rate for different topologies is given by SU(3)
colour rules in combination with a minimization of the $\lambda$ measure.
The many junctions leads to an enhanced baryon production, although
partly compensated by a shift towards strings with masses too low for
baryon production.

A specific application of CR is for $t$, $Z^0$ and $W^{\pm}$ decays. With
widths around 2~GeV, \textit{i.e.} $c\tau \approx 0.1$~GeV, their decays
happen after other hard perturbative activity (ISR/FSR/MPI) but still
inside the hadronization colour fields, thereby allowing CR with the rest
of the event. It was already for the Tevatron noted that this is a
non-negligible source of uncertainty in top mass determinations
\cite{Skands:2007zg}, and for similar LHC studies several new CR models
were implemented in  \textsc{Pythia}~8 \cite{Argyropoulos:2014zoa}.
These fall in two classes: either the $t$ and $W$ decay products undergo
CR on equal footing with the rest of the event, or their decays and CR
are considered after the rest of the partonic event has had a chance to
reconnect. The latter scenario allows more flexibility, to explore also
extreme possibilities.

\section{LHC lessons}
\label{tssec:lessons}

LHC has been very productive in presenting data of relevance for
MB/UE/MPI/DPS studies, and there is no possibility to cover even a
fraction of it here. The following therefore is a very subjective
selection.

To begin with, some words on tunes, see also
\cite{CONTRIB:EXP-UE,CONTRIB:MC-TUNING-PROFESSOR}.
Generators contain a large number
of free parameters, by necessity, that attempt to parametrize our
ignorance. Many of these are correlated, so cannot be determined
separately. A tune is then the outcome of an effort to determine
a set of key parameters simultaneously. This is an activity that
generator authors do at a basic level all the time, and occasionally
as a more concerted effort, \textit{e.g.}
\cite{Skands:2010ak,Corke:2010yf,Schulz:2011qy,Skands:2014pea}.
It is also an activity that experimental collaborations undertake,
given the direct access to data and the needs of their communities.
With no claims of completeness, the \textsc{Pythia}~6 code contains
settings for more than 100 different tunes, \textit{e.g.}
\cite{Buttar:2004iy,Field:2005sa,Field:2006iy,ATLAS:2010zyu,%
ATLAS:2011zja}, whereof about half precede the LHC startup,
and the \textsc{Pythia}~8 one for 34 so far, \textit{e.g.}
\cite{ATLAS:2012uec,ATLAS:2014xxx,Aad:2014xaa,Khachatryan:2015pea}.
Many of these are minor variations around a common theme. Some tunes have
been made by hand whereas others use automated procedures such as
\textsc{Professor} \cite{Buckley:2009bj}. The access to validated
\textsc{Rivet} analysis routines \cite{Buckley:2010ar} have played an
increasingly important role. Data comparisons for many tunes are
available in \textsc{MCplots} \cite{Karneyeu:2013aha}.

Whereas theorists aim for the best overall description, experimentalists
often produce separate MB and UE tunes. With less data to fit, it is
possible to obtain a better description for the intended applications.
So far it has not been settled how much of the differences in MB and UE
parameters represent true shortcomings of the model and how much is
a consequence of the fitting process itself.

Several \textsc{Pythia}~6 tunes served as a basis for predictions
prior to the LHC startup. Early 7~TeV data  showed that
most of them undershot the level of activity by
some amount, with one being close, but none above. Fortunately a modest
change of the energy dependence of the $p_{\perp 0}$ is enough to bring up
the activity to a reasonable level, and further extrapolations to 13~TeV
have worked better.

Generally speaking, \textsc{Pythia} has been able to
explain most phenomenology observed at LHC so far, at least qualitatively,
and often also quantitatively. Nevertheless, a significant number of
observations do not look as nice. A common theme for many of them is that
high-multiplicity $pp$ events have properties similar to those observed
in heavy-ion $AA$ collisions. Examples are\cite{Fischer:2016zzs}
\begin{Itemize}
\item High-multiplicity events have a higher fraction of heavier particles,
notably multi-strange baryons \cite{ALICE:2017jyt},
whereas the composition stays rather constant in \textsc{Pythia}.
\item The $\langle p_{\perp} \rangle$ is larger for heavier particles
\cite{Abelev:2014qqa} (also observed at RHIC \cite{Abelev:2006cs}),
more so than \textsc{Pythia} predicts.
\item The charged particle $p_{\perp}$ spectrum is underestimated
at low $p_{\perp}$ scales \cite{Aad:2010ac,Chatrchyan:2011av,Adam:2015pza},
\textit{e.g.}\ leading to problems in simultaneously fitting MB data
analyzed with $p_{\perp} > 0.1$~GeV and $p_{\perp} > 0.5$~GeV. The deficit is
mainly associated with too little $\pi^{\pm}$ production \cite{Adam:2015qaa}.
\item The $\Lambda / K$ spectrum ratio has a characteristic peak at
around 2.5~GeV \cite{Khachatryan:2011tm}, which is not at all reproduced.
\item The observation of a ridge in two-particle correlations, stretching
out in rapidity on both sides of a jet peak, especially for high-multiplicity
events \cite{Khachatryan:2010gv,Aad:2015gqa,Khachatryan:2016txc}.
Correlation functions also points to azimuthal flow, similarly to
$AA$ observations.
\end{Itemize}

An alternative model has been formulated \cite{Fischer:2016zzs} to explore
at least some of these discrepancies, with three main deviations from the
standard framework. Firstly the standard Gaussian $p_{\perp}$ spectrum for
primary hadron production is replaced by an exponential one,
$\exp( - m_{\perp \mathrm{had}} / T)$. It is loosely inspired by thermodynamics,
with $T$ an effective temperature, and is intended to enhance the pion
rate at small $p_{\perp}$, while increasing $\langle p_{\perp} \rangle$
for heavier particles. Secondly, it is assumed that the normal string
tension, alternatively the $T$ above, is increased in regions of phase
space where strings are close-packed, which typically is caused by a
high MPI activity. This is intended to change both particle composition
and $p_{\perp}$ spectra. Thirdly, a simple model for hadronic rescattering
is introduced, whereby hadrons tend to obtain more equal velocities,
\textit{i.e.}\ larger $\langle p_{\perp} \rangle$ for heavier particles.

Unfortunately, effects are not as large as one might hope. Specifically,
most pions come from decays of heavier hadrons, and so the mechanisms
intended to give less $p_{\perp}$ to pions and more to kaons and protons
are counteracted. Nevertheless the thermodynamical model is able to
provide significantly improved descriptions of observables such as the
$p_{\perp}$ spectrum of charged hadrons, the average transverse momentum
as a function of the hadron mass, and the enhanced production of strange
hadrons at large multiplicities.

Even more successful are the \textsc{Dipsy} and \textsc{Epos} generators,
however. In \textsc{Dipsy} dense string packing is assumed to lead to the
formation of colour ropes \cite {Bierlich:2014xba}, wherein an increased
string tension favours the production of heavier hadrons and larger
$p_{\perp}$ values, and a shoving mechanism can induce ridge and related
phenomena \cite{Bierlich:2016vgw}. In \textsc{Epos} the central region
of $pp$ collisions can form a quark--gluon plasma,
which also allows strangeness enhancements, and strings in the outer
regions can again be shoved by the central pressure, to give ridges
\cite{Pierog:2013ria}.

These new data, and the models they favour, may have consequences for
the way we think about MPIs. Having MPIs as the origin of QGP formation
in $AA$ is an old idea \cite{Eskola:1988yh}, but now it might even apply
to $pp$. One could also note that the rising trend of
$\langle p_{\perp} \rangle (n_{\mathrm{ch}})$, once the key reason to
introduce CR, now partly might be attributed to collective effects.

\section{Current state of the PYTHIA MPI machinery}
\label{tssec:current}

The MPI machinery implemented in \textsc{Pythia} has evolved over
the years, as we have seen. It is therefore useful to make a quick
rundown of the current state, and also mention a few
odds and ends that were not yet covered.

The starting point is to define a
$\mathrm{d} \sigma / \mathrm{d} p_{\perp}^2$, Eq.~(\ref{tseq:dsigma}),
that decides which processes can occur in an MPI, and then also occurs
in the Sudakov-like factor. Originally it only contained QCD $2 \to 2$
processes, but now also includes $2 \to 2$ with photons in the final
state, or mediated by an $s$-channel $\gamma^*$, or by $t$-channel
$\gamma^*/Z^0/W^{\pm}$ exchange, or charmonium and bottomonium production
via colour singlet and octet channels. This combined cross section is
then regularized in the $p_{\perp} \to 0$ limit by the damping factor in
Eq.~(\ref{tseq:ptdamp}).

There are two main options to begin the generation. One is if a hard
process already has been selected, with a generic hardness scale,
\textit{e.g.}\ the factorization scale, which we for the sake of bookkeeping
equate with $p_{\perp 1}$. Then the impact parameter $b$ can be selected
according to one of five possibilities: no $b$ dependence, single Gaussian,
double Gaussian, $\exp (-b^p)$ overlap and $x$-dependent Gaussian. The
other main option is \textit{e.g.}\ for inclusive non-diffractive production,
where $p_{\perp 1}$ and $b$ have to be selected correlated according to
Eq.~(\ref{tseq:phardestwithb}). (Also with an  explicit $x_1, x_2$
dependence for the $x$-dependent Gaussian.)

The downwards evolution in $p_{\perp}$ can then proceed,
Eq.~(\ref{tseq:pnextwithb}).
For a preselected hard process there is an ambiguity, however,
whether to allow $p_{\perp 2}$ to be restricted by $p_{\perp 1}$ or go
all the way up to the kinematic limit. The former is required for
QCD jets, to avoid double-counting, while the latter would be sensible
for more exotic processes, where there is no such risk. By default
\textsc{Pythia} will make this decision based on some simple rules,
but it is also possible to choose strategy explicitly.

The downwards evolution of MPIs is interleaved in $p_{\perp}$ with
ISR and FSR, Eq.~(\ref{tseq:combinedevol}). Optionally one may also
include rescattering in the MPI framework, Eq.~(\ref{tseq:combinedresc}),
but this comes at a cost, so is not recommended for normal usage.
Joined interactions are not implemented in the \textsc{Pythia}~8 code,
and there is also no swing mechanism.

Modified multiparton PDFs during the evolution follow the strategy of
Sec.~\ref{tssec:multiPDFs}. The same section also describes the related
handling of beam remnants. The subsequent colour reconnection stage
allows for several different models, with many options. String
fragmentation and decays is added at the end of the generation chain,
with junction topologies playing a key role for the preservation of
the incoming baryon numbers.

Diffractive topologies are included in a picture where the Pomeron
is given a hadronic substructure, implying that $\mathbb{P}p$ and
$\mathbb{P}\mathbb{P}$ subcollisions should be handled with a full MPI
machinery of its own, at least for higher diffractive masses.

In total, essentially all the questions raised about the original model,
end of Sec.~\ref{tssec:firstmodel}, have since been studied, and
tentative answers have been implemented in the existing code.

The generation of DPS is implicit in the MPI machinery. Whereas the
first interaction can always be selected hard, normally the second one
would tend to be soft. There is a special machinery in \textsc{Pythia}
to generate two hard interactions. To understand how it operates,
start from the Poissonian distribution
$\mathcal{P}_n = \langle n \rangle^n \, \exp( - \langle n \rangle) / n!$.
If $ \langle n \rangle \ll 1$, as it should be for a hard process,
the exponential can be neglected and
$\mathcal{P}_2 = \langle n \rangle^2 / 2$. Now imagine two processes
$a$ and $b$ with cross sections $\sigma_a$ and $\sigma_b$, meaning that
they are produced with rates
$\langle n_a \rangle = \sigma_a / \sigma_{\mathrm{nd}}$ and
$\langle n_b \rangle = \sigma_b / \sigma_{\mathrm{nd}}$ inside the
inelastic event class. Then the cross section for having two such MPIs is
\begin{equation}
\sigma_2 = \sigma_{\mathrm{nd}} \, \mathcal{P}_2
= \sigma_{\mathrm{nd}} \,
\frac{(\langle n_a \rangle + \langle n_b \rangle)^2}{2!}
= \frac{\sigma_a^2 + 2 \sigma_a \sigma_b + \sigma_b^2}{2 \, \sigma_{\mathrm{nd}}}
\label{tseq:dpsmaster}
\end{equation}

The above equation neglects the impact-parameter dependence. A hard
collision implies a smaller average $b$ than for MB events, as we have
discussed before, and thus an enhanced rate for the second collision.
Poissonian statistics applies for a fixed $b$, but when averaging over
all $b$ an enhancement factor $\mathcal{E}$ is generated. Conventionally
such effects are included by replacing the final $\sigma_{\mathrm{nd}}$
in Eq.~(\ref{tseq:dpsmaster}) by an $\sigma_{\mathrm{eff}}$. Unintuitively
a lower $\sigma_{\mathrm{eff}}$ corresponds to a higher $\mathcal{E}$.
It can be written as \cite{Seymour:2013sya}
\begin{equation}
\mathcal{E} = \frac{\sigma_{\mathrm{nd}}}{\sigma_{\mathrm{eff}}}
= \frac{\int \widetilde{\mathcal{O}}^2(b) \, \mathrm{d}^2b \times%
\int \mathcal{P}_{\mathrm{int}}(b) \, \mathrm{d}^2b}%
{\left( \int \widetilde{\mathcal{O}}(b) \, \mathrm{d}^2b \right)^2} ~.
\label{tseq:enhance}
\end{equation}
Thus $\mathcal{E}$ depends on the shape of $\widetilde{\mathcal{O}}(b)$,
with a distribution more spiked at $b = 0$ giving a larger
$\mathcal{E}$. But $\mathcal{E}$ also depends on the CM energy and the 
$p_{\perp 0}$ scale, which enter via $\mathcal{P}_{\mathrm{int}}(b)$.

There is also a dynamical depletion factor related to PDF weights.
With the two hard processes initially generated independently of each
other, flavour and momentum constraints are not taken into account.
Therefore, afterwards, PDFs are re-evaluated as if either interaction
were the first one, giving modified PDFs for the second one, as described
in Sec.~\ref{tssec:multiPDFs}. The average PDF weight change of the two
orderings is used as extra event weight, leading to some configurations
being rejected.

The program allows the two hard interactions to be selected in partly
overlapping channels, and/or (with some warnings) phase space regions.
Assume \textit{e.g.}\ that process~1 can be either $a$ or $c$ and
process~2 either $b$ or $c$. Then an extension of
Eq.~(\ref{tseq:dpsmaster}) tells us the numerator should be
\begin{equation}
2 \sigma_{1a}\sigma_{2b} + 2 \sigma_{1a}\sigma_{2c} +
2 \sigma_{1c}\sigma_{2b} + \sigma_{1c}\sigma_{2c}
= 2 (\sigma_{1a} + \sigma_{1c})(\sigma_{2b} + \sigma_{2c})
- \sigma_{1c} \sigma_{2c}
\end{equation}
To obtain the correct answer the prescription is thus to generate
process 1 according to $\sigma_a + \sigma_c$ and process 2 according
to $\sigma_b + \sigma_c$, but to throw half of those events where
both 1 and 2 were picked to be process $c$.

\section{Summary and Outlook}
\label{tssec:summary}

In this chapter we have traced the evolution of MPI ideas in
\textsc{Pythia} over more than 30 years. The emphasis has been on
the early developments, since that set the stage for what has come
later, and more generally on the theoretical ideas and concepts that
have been explored over the years. It is intended to offer a
complementary view to other chapters in this book, and so we have
avoided detailed comparisons with data, and also not gone into
details of other models.

By and large, the original MPI ideas have stood the test of time;
they are still at the core of the current \textsc{Pythia} framework.
The further work that has been done since is much more extensive
than the original one, but often suffers from diminishing returns,
\textit{i.e.}\ that major upgrades only moderately improve the general
agreement with data. Nevertheless it is important to explore as many
aspects of MPIs as possible. They do encode important information on the
borderline between perturbative and nonpeturbative physics, and we
should become better at decoding this information.

The \textsc{Pythia} development described here has been very much
influenced by perceived experimental needs, and inspired by theoretical
ideas, but has been decoupled from detailed theoretical calculations.
The reason is obvious: already DPS offers a formidable challenge,
enough to keep theorists busy, and so useful results for truly
\textit{multi}parton interactions are rare. While it is interesting
to understand two-parton PDFs better, say, in \textsc{Pythia} we need
to be able to address twenty-parton PDFs, and nothing less will do.

In this spirit it is important to recall that, even though studies of
DPS is an important way to explore MPI, it is not the only one. An
example on the to-do list is the lumpiness of particle production in
general, \textit{e.g.}\ as probed by the minijet rate for different
jet clustering $R$ and $p_{\perp\mathrm{min}}$ parameter values, down to
the $p_{\perp 0} \approx 2.5$~GeV scale (at LHC energies). And it should
not be forgotten that (for some people, like me) the most convincing
--- and earliest --- evidence of MPIs is the broadness of multiplicity
distributions. The intermittency \cite{Bialas:1985jb,Bialas:1988wc}
interpretations of multiplicity fluctuations at the S$p\overline{p}$S
\cite{Albajar:1990wp,DeWolf:1995nyp} may have been over-enthusiastic
at times, but more parts of the same experimental program could be
carried out at the LHC and yield useful results.

Nevertheless, the LHC has provided new impetus to the whole MB/UE field
of studies, by the observation of new and unexpected phenomena. Both
the ridge effect and the enhancement of multi-strange baryons in
high-multiplicity events, to take the two most spectacular examples,
remain to be fully understood. These observations do not invalidate
the MPI concept. On the contrary, plausible explanations start out
from a MPI picture and add some kind of collective behaviour among
the MPIs. Colour ropes or other ways to obtain an increased string
tension is one example, the formation of a quark-gluon-plasma-like
(multiparton!) state another.

Clearly much work lies ahead of us to fully understand what has
already been observed, and hopefully also many further surprises
will come along to stimulate us further.

\section*{Acknowledgements}

Thanks to all collaborators on MPI physics through the years, and to
Jonathan Gaunt and Peter Skands for helpful comments on the draft
manuscript. This project has received funding in part by the Swedish
Research Council, contracts number 621-2013-4287 and 2016-05996,
in part from the European Research Council (ERC) under the
European Union's Horizon 2020 research and innovation programme
(grant agreement No 668679), and in part through the European
Union Marie Curie Initial Training Networks 
MCnetITN PITN-GA-2012-315877 and MCnetITN3 722104.

\bibliographystyle{utphys}
\bibliography{lutp1722}

\end{document}